\documentclass[journal]{IEEEtran}
\ifCLASSINFOpdf

\else

\fi
\usepackage{mathrsfs}

\usepackage{amsmath}
\usepackage{ntheorem}

\usepackage{makeidx}  
\usepackage{algorithm}
\usepackage{algorithmic}

 \usepackage{graphicx}

\usepackage{subfigure}
\usepackage{epstopdf}
\usepackage{bm}
\usepackage{cite}
\usepackage{stfloats}
\usepackage{setspace}

\usepackage{amssymb}
\usepackage{url}
\newtheorem{proposition}{Proposition}
\usepackage{tabularx,booktabs}
\newcolumntype{C}{>{\centering\arraybackslash}X} 
\usepackage{lipsum}

\usepackage{makecell} 

\usepackage{color}

\newtheorem{remark}{Remark}

\usepackage{multirow}

\definecolor{dblue}{RGB}{15,89,164}
\usepackage{amsmath,amssymb}

\begin{document}
\title {Near-Field Communications with Different Array Geometries: Rayleigh Distance, Channel Estimation, and Transmission Design\vspace{-10pt}}
\author{ Kangda Zhi,  Yi Song, Tianyu Yang, Tuo Wu, Tengjiao Wang, and Giuseppe Caire, {\itshape Fellow, IEEE \upshape}		
		\thanks{ K. Zhi is with Guangdong Provincial Key Laboratory of Aerospace Communication and Networking Technology, Harbin Institute of Technology (Shenzhen), Shenzhen, China, 518055 (zhikangda@hit.edu.cn)
			Y. Song,  T. Yang, and G. Caire are with Technische Universit\"{a}t Berlin, Germany (\{k.zhi,   yi.song, caire\}@tu-berlin.de). Tuo Wu is with South China University of Technolog, China (wutuo@scut.edu.cn).}
	  \vspace{-35pt}
}

\maketitle
\begin{abstract}
 This work establishes a framework of near-field communication under different array geometries for extremely large-scale multiple-input multiple-output (XL-MIMO). We first formulate the near-field spatially non-stationary channel model which is characterized by the distance between the user and each antenna on uniform and modular curved arrays. By fixing the total number of antennas while varying the degree of curvature, we investigate a fair case where the horizontal arc length of the curved array is the same as the planar array.  We explicitly unveil the non-trivial impact of array curvature on extending the near-field region for cell edges.  Then, for arbitrary array geometries and arbitrary-field channels, we estimate the spatial-domain channel by tackling a compressed sensing problem with a learned regularizer. Without relying on specific codebooks, we propose a denoising autoencoder (AE)-aided approximated message passing (AMP) algorithm and provide the corresponding theoretical replica bound. Finally, based on the estimated channel, we propose an optimization algorithm to maximize the sum user rate for sub-connected XL-MIMO systems by jointly designing the array geometry and hybrid precoding in the downlink. Numerical results demonstrate that the proposed AE-AMP algorithm can effectively estimate the spatially non-stationary near-field channels with robustness and generality compared to several conventional and deep-learning-based benchmarks. The improvement of data rate by using modular curved arrays with the estimated channel is also validated.
\end{abstract}


\IEEEpeerreviewmaketitle

 \vspace{-10pt}
\section{Introduction} \label{section0}
To meet the ever-increasing data rate requirement for the upcoming sixth generation (6G) network, a key approach consists of substantially increasing the number of antennas, leading to the emergency of extremely large-scale multiple-input multiple-output (XL-MIMO)\cite{Lu2024Survey,mao2025neft,yang2024near}. By incorporating hundreds and even thousands of antennas, XL-MIMO is expected to achieve order-of-magnitude higher spectral efficiency than conventional massive MIMO systems. In addition,  the enlarged array aperture is helpful to achieve high spatial resolution, which is beneficial for accurate channel estimation, sensing, and wireless localization services in 6G systems.

The deployment of XL-MIMO  introduces new challenges. With a large number of antennas,  it is highly possible that  users or scatterers will be located in the near field of the XL-MIMO, since  the Rayleigh distance is proportional to the square of the array aperture\cite{zhi2024performance}.  In this case, propagation by spherical waves should be considered, making the conventional channel modeling based on far-field planar wave assumption invalid. Near-field propagation also exhibits some other unique features, such as  beamfocusing to tackle co-angle interference\cite{chen2025unified}, spatial non-stationarities\cite{tang2026tutorial,liu2026TSP} and hybrid near- and far-field properties\cite{zhang2026rotatable} which distinguish the research of XL-MIMO from conventional MIMO foundations.

To enable  high-quality data transmission in XL-MIMO systems, several challenges should be tackled. The first fundamental issue is how to obtain the  channel state information (CSI) under near-field and even hybrid-field propagations, in an accurate, general, and low-overhead way. Given the spherical-wavefront properties, existing literature has investigated the design of angle-distance-domain near-field codebooks by guaranteeing that the coherence between any two codebook vectors is below a desired threshold, with respect to uniform linear arrays (ULAs) \cite{cui2022channel},  uniform planar arrays (UPAs) \cite{wu2023multiple}, and uniform circular arrays\cite{wu2023enabling}.  Meanwhile,  several  methods have been proposed to reduce the overhead in parameter estimation and beam training,  such as using the idea of  sub-array decomposition\cite{zhi2025localization}, exploring features of far-field beam\cite{wu2023two}, using hierarchical codebooks \cite{zhang2023codebook}, and exploiting deep learning to find the optimal codewords\cite{liu2022deep,wang2026robust}.    Besides, some researchers have studied the estimation and tracking problem under hybrid-field  and  spatial non-stationary channels\cite{tang2024joint,xu2024joint}.

Assuming that the near-field CSI is perfectly known, another important challenges raised by XL-MIMO architecture is how to design an efficient transmission scheme and understand the inherent mechanism. To this end, the asymptotic limit of near-field communication was revealed in \cite{zhi2024performance,lu2021communicating} and the conventional Rayleigh distances have been refined to consider the impact of incident angle and misalignments\cite{wu2023enabling,zhang2026near}. The properties of near-field beam depth and degree of freedom have also been analysed in\cite{kosasih2024finite,ouyang2023near}. Meanwhile, the optimization design of near-field beamfocusing has been widely investigated across diverse scenarios including physical security\cite{zhang2024physical}, wireless sensing\cite{wang2023near},  and multiple access\cite{zhang2025rate}. Furthermore, the additional gain by adjusting the array topologies and geometries on near-field communications have been explored, based on the architecture of fluid and movable antennas\cite{chen2024joint,zhu2025movNF,liu2026movNF}, non-uniform array\cite{guo2025design}, modular and sparse array\cite{shen2025hybrid,wu2026beam}.

In the presence of the above-mentioned contributions, there still exist research gaps in enhancing the reliability and  quality of near-field communication. From the perspective of channel estimation, firstly, the existing methods commonly rely on specific array geometries such as ULA, UPA, and uniform circular array, to exploit the mathematical structure of steering vectors and design specific codebooks\cite{cui2022channel,wu2023multiple,wu2023enabling}. As the shape and topology of arrays change, the near-field codebook needs to be redesigned, which is not favourable in practice. This challenge exacerbates in the presence of hybrid-field channels and spatial non-stationarity, due to more complicated channel structures. Secondly,  although efforts have been made to reduce the estimation and training overhead, the  codebook-based or grid-based methods suffer from inherent limitations of high codebook size and grid mismatch in XL-MIMO systems, due to the extremely large number of antennas and the high searching dimension for angle-distance-delay-domain parameters, especially when the transmitter and receiver both have multiple antennas\cite{zhi2025localization}. Therefore, it is of practical value to investigate the general channel estimation method for XL-MIMO systems without relying on specific assumptions of channel structures and codebooks and with low complexities.

Given the general channel estimation method, next, it would be possible to investigate the practical performance of transmission schemes for XL-MIMO under the design of precoding and array geometries. Although existing research has investigated the uniform circular and cylindrical arrays and shown their advantages compared to linear and planar arrays, the comparison commonly assumes the same aperture between them. In this case, a curved array would have a larger number of antennas due to the bent shape and therefore yield superior performance, unsurprisingly. This motivates a natural question: given a service area in the cell, what will happen if a fixed-length linear/planar array is bent to a certain angle, and what is the change of near-field boundaries and near-field precoding performance? This needs new efforts for both theoretical analysis and optimization design. To the best of the author's knowledge, the joint design of precoding and array geometry, along with the enabling channel estimation method, has not been investigated yet, which, however, is an important issue in practice.

Motivated by the above research gaps,   this paper aims to establish an analysis and optimization framework of XL-MIMO systems in the presence of near-field spatial non-stationary channels and with different array geometries. We  first formulate the general channel model given the number of antennas at different array curvatures and shapes and then propose  deep learning-based channel estimation methods without relying on specific dictionaries/codebooks. We also analyse the impact of array curvature on near-field distances and optimize the corresponding transmission schemes based on the estimated channels. The specific contributions are summarized as follows.

 		\begin{itemize}
\item  Given the fixed number of antennas, we establish the general spatial non-stationary channel model for XL-MIMO where the uniform linear and planar array can be bent to a certain curvature. We derive the theoretical and effective Rayleigh distances and analyse the impact of array curvature on near-field boundaries. With a fixed arc length, we demonstrate that the Rayleigh distance increases laterally while decreasing normally  as the degree of array curvature increases, which unveils that there is an optimal array curvature to maximize the near-field area. Finally, the model of uniform curved array is extended to  modular cylindrical arrays (MCAs)  with sparse subarrays for a larger near-field region.

\item To estimate the channel under arbitrary array geometries and arbitrary fields, we proposes  a deep learning-based   approximate message passing (AMP) algorithm which uses a denosing autoencoder (AE) to learn the low-dimensional channel manifold and serves as a learned regularizer for the compressed sensing problem. We also establish the state evolution (SE) and replica bound of the proposed AE-AMP algorithm to evaluate its theoretical performance. Several benchmarks are also proposed for comprehensive comparisons.

\item We formulate the problem to maximize the sum user rate with  joint optimization of array geometries and hybrid precoding. The curvature angle, subarray number, and subarray spacing of MCA are designed to minimize the average channel condition number so that they do not need to be adjusted with time. Meanwhile, effective algorithms are proposed to establish closed-form solutions for designing digital and analog precoders in each iteration.

\item Numerical results are provided to validate the  insights of adjusting array geometries, the accuracy of the channel estimation algorithms, and the effectiveness of the optimization framework.

\end{itemize}


\section{System Model}\label{section2}

\begin{figure}
	\centering
	\includegraphics[width= 0.48\textwidth]{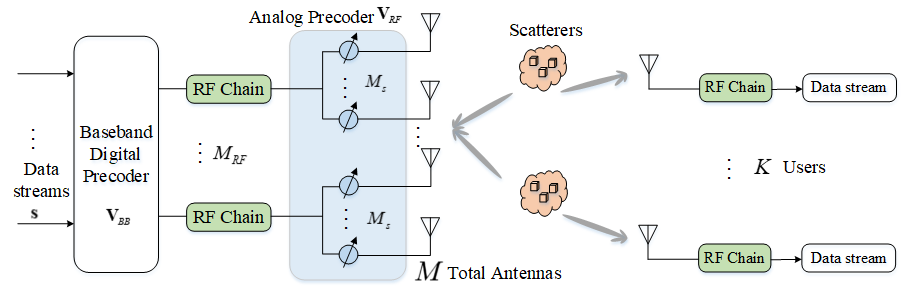}
	\vspace{-15pt}
	\caption{Illustration of the considered sub-connected XL-MIMO system.}
	\vspace{-15pt}
	\label{figure1}
\end{figure}

As shown in Fig. \ref{figure1}, a multi-antenna base station (BS) with a hybrid structure is considered to communicate with $K$ single-antenna users. The BS is equipped with $M$ antennas and $M_{RF}$ radio-frequency (RF) chains, where $M_{RF}\ll  M$. Employing a sub-connected structure, each RF chain at BS is connected to $M_s=M/M_{RF}$ antennas. In general, it is assumed that the antennas are arranged as patches on a surface, which degenerates to a plane for two-dimensional (2D) case (e.g., UPA) and to a line for one-dimensional (1D) case (e.g., ULA). The specific enumeration of the antenna elements will be specified for each individual geometry later.


Near-field spatial non-stationary channels are considered given the large number of antennas at the BS.  The  propagation between user $k$  and the BS is modeled as a geometric channel with one line-of-sight (LoS) path and $L_k$ non-LoS (NLoS) paths.  As in \cite{han2020channel,liu2026TSP},  the   spatial non-stationarity of each path is characterized by a visibility region (VR) set $ \Phi_{k,l} $, $0\leq l \leq L_k$,  which  contains the indices of antennas over which path $l$ of user $k$ is visible.  Accordingly, the  channel from user $k$ to the BS, denoted by $\mathbf{h}_k\in \mathbb{C}^{M \times 1}$,   is expressed as
\begin{align}\label{channel_H}
	\begin{aligned}
		\mathbf{h}_k   =  \underbrace{{\mathbf{h}}_{k,0}  \odot 
		\mathbf{p}\left(\Phi_{k,0}\right)}_{\mathbf{h}_k ^{\rm los} }
		+ \underbrace{\sum\nolimits_{l=1}^{L_k} A_{k,l}\alpha_{k,l}
			{\mathbf{h}}_{k,l}   
			\odot 
			\mathbf{p}\left(\Phi_{k,l}\right)  }_{ \mathbf{h}_k ^{\rm nlos} },
	\end{aligned}
\end{align}
where $\odot$ denotes the Hadamard product. The vector $\mathbf h_{k,l}$ denotes the array response associated with path $l$ before applying the VR mask.  $\mathbf p(\Phi_{k,l})\in \{0,1\}^{M\times 1}$ is an antenna-level binary visibility vector associated with the  VR set $\Phi_{k,l}$. Specifically, if  the index of antenna $m$   is   included in $\Phi_{k,l}$, then $[\mathbf p(\Phi_{k,l})]_m=1$; otherwise, $[\mathbf p(\Phi_{k,l})]_m=0$.

The $m$-th element of $  {\mathbf{h}}_{k,0}  $  in (\ref{channel_H}) denotes the near-field LoS channel  between    user $k$  and the $m$-th antenna of the BS, which can be modeled by\cite{Lu2024Survey}
\begin{align}\label{channel_los}
  \left[{\mathbf{h}}_{k,0}\right]_m  =  {  ( {U_m}}/{r_{k,0}^{(m)}}  )   \exp \left({-j {2\pi} r_{k,0}^{(m)}}    /{\lambda}   \right),
\end{align}
where   $r_{k,0}^{(m)}$ denotes the distance between the antenna of user $k$  and the $m$-th antenna of the BS.  $U_m$ denotes the amplitude-domain radiation factor of the $m$-th antenna along the propagation direction.    $ 	{\mathbf{h}}_{k,l}    $ in (\ref{channel_H})  with $l\in [1,L_k]$ correspond to the  NLoS paths associated with the $l$-th scatterer. Its  $m$-th element is characterized by the distance $r_{k,l}^{(m)}$ from the $l$-th  scatterer to the $m$-th antenna of the BS array, in a manner similar to (\ref{channel_los}). 
  $A_{k,l}$ and $\alpha_{k,l} \sim \mathcal{C N}(0,1)$ denote the additional path gain and random small-scale fading coefficient associated with the signal scattered by the $l$-th scatterer from user $k$, respectively.

\begin{remark}
	Channel  (\ref{channel_H}) is of a general form and the detailed expression of $ \mathbf{h}_k   $ relies on  specific calculation of distances $r_{k,0}^{(m)}$ and $r_{k,l}^{(m)}$, $\forall m$, which depends on array geometries.
In near fields, these distance parameters are a function of angle of arrivals (AoAs) and the distance from the user to the center of BS array.    When users move far away from the BS,  based on the accuracy of linear Taylor approximation of distances associated with different paths\cite{zhi2024performance},  near-field channels can be gradually  approximated by mixed near- and far-field channels and fully far-field channels.
Besides, the next section will model the channel of XL-MIMO with different array geometries by characterizing the distances with additional hardware-related parameters, including array curvature angle, number of subarrays, and subarray spacing.
\end{remark}

\vspace{-15pt}
\section{Channel Modeling  and Analysis with Different Array Geometry}\label{section3}
Existing studies typically model curved arrays as a full circles or cylinder with the same effective aperture as  linear/planar arrays. 
This approach is unsuitable for a fair comparison between curved and planar arrays, since a full circular/cylinder array inevitably contains far more antennas than its linear/planar counterpart given the same aperture. Moreover, it precludes any study of how the array curvature reshapes the near-field region. To enable a fair comparison, in this section we fix the number of antennas and treat the array curvature as a tunable parameter\footnote{For brevity, we only present the results for LoS path and with a single user. The modelling of NLoS path and for multiple users can be readily done in the same way.}.

\subsection{Uniform Curved Array}
\begin{figure}[t]
	\centering
	\includegraphics[width= 0.45\textwidth]{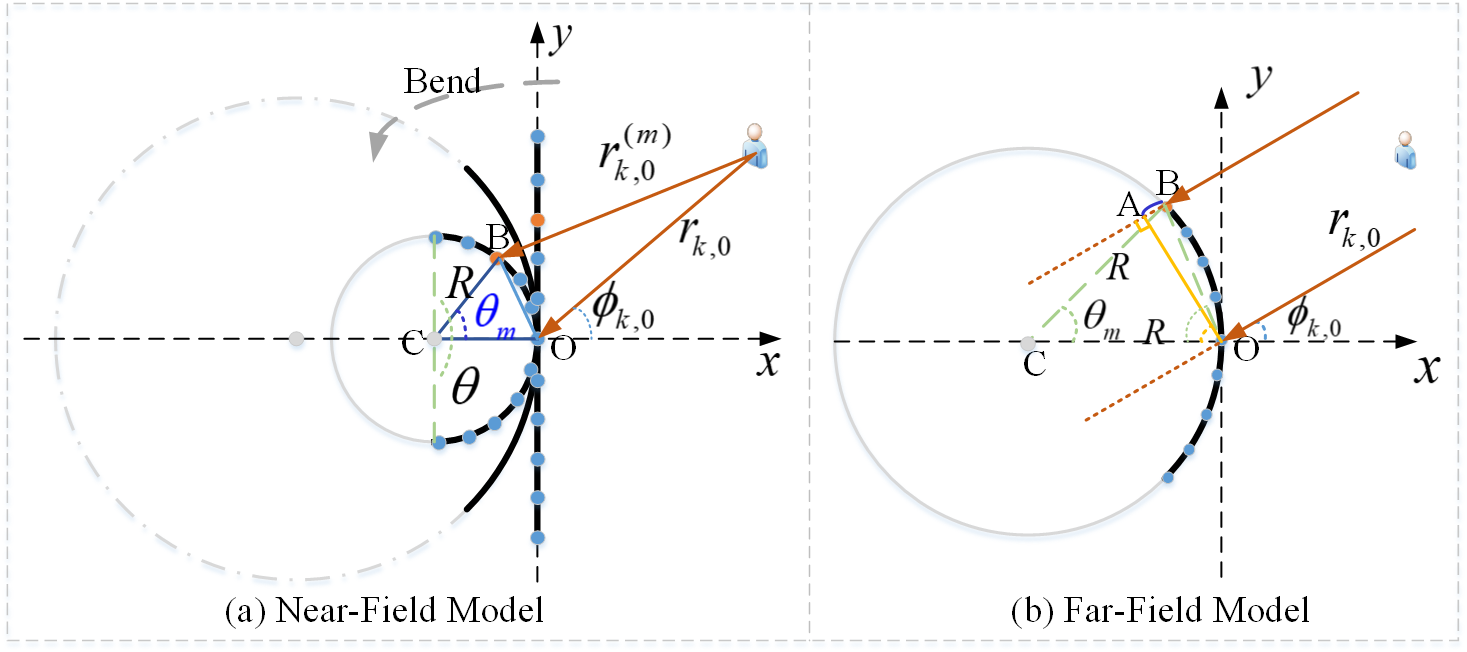}
			\vspace{-10pt}
	\caption{The change of array curvature angle $\theta$ given fixed curve length $L$.}
			\vspace{-15pt}
			\label{figure11}
\end{figure}
To begin with, we consider the 2D uniform curved array for analytical tractability and clarity. Specifically, consider a uniform linear array with $M$ antennas and with antenna spacing $d$ and array length   $L=(M-1)d$. Now, fix the array length $L$ (i.e., fixing $M$), but change its curvature, as shown in Fig. \ref{figure11}. Assume that the curvature angle is $\theta$  (rad). Then, the curved array is a part of circular array with a radius of $R = \frac{L}{\theta}$. Intuitively, it can be seen that, as the array curvature angle increases, the array bends more, thus reducing its projected aperture in front of the array but extending that in the sides. 

 Let the center of the ULA be the origin of the coordinate system, i.e, point $O$ in Fig. \ref{figure11}. Define the incident angle of the spherical wave from user $k$ to the ULA as $\phi_{k,0}$. Define $r_{k,0}$  as the distance from user $k$ to the central antenna. Then, the angle $\angle BCO$ (see Fig. \ref{figure11}) of the $m$-th antenna is given by $\theta_m = m \frac{\theta}{M-1}$, $m=-\frac{M-1}{2} , \cdots, \frac{M-1}{2}$. 


Let the location of the user $k$ and the $m$-th antenna in the 2D space be denoted by  $\mathbf{u}_k= (r_{k,0} \cos\phi_{k,0}, r_{k,0} \sin\phi_{k,0})$ and  $\mathbf{p}_m = (R \left(\cos\theta_m - 1\right), R\sin\theta_m)$, respectively. Then, the distance between the user and the $m$-th antenna is given by
\begin{align}\label{fsdgsdfgsdfhgsdgh}
	\begin{aligned}
		r_{k,0}^{(m)} &\!=\!    \left\|\mathbf{u}_k \!-\!  \mathbf{p}_m\right\| \!=\! 
		\big\{\left[r_{k,0} \cos \phi_{k,0}-R\left(\cos \theta_m-1\right)\right]^2\\
		&+\left(r_{k,0} \sin \phi_{k,0}-R \sin \theta_m\right)^2  \big\}^{1/2} \\
		& = {r_{k,0}}    \big\{ 1 +    \left[    { 2R\sin\left({\theta_m}/{2}\right)   }/{r_{k,0}} \right]^2  - \\
		&    {4R}{} \sin \left({\theta_m}/{2}\right) \sin  \left(\phi_{k,0} - {\theta_m }/{2}   \right) /  r_{k,0}   \big\}^{1/2}.
	\end{aligned}
\end{align}
Using the first and second order Taylor expansion of (\ref{fsdgsdfgsdfhgsdgh}), we obtain the expressions of the distance for the far-field and near-field channels in the form of
\begin{align}\label{fist_order_taylor_UCA}
	&r_{k,0,\rm far}^{(m)}  \approx r_{k,0}-  {2 R} \sin \left({\theta_m}/{2}\right) \sin \left(\phi_{k,0}- {\theta_m}/{2}\right) ,\\ \label{second_order_taylor_UCA}
	&r_{k,0, \rm near}^{(m)} \!  \approx \!  r_{k,0,\rm far}^{(m)}  \! +\!   \frac{2 R^2}{r_{k,0}} \sin ^2\left({\theta_m}/{2}\right) \cos ^2 \!  \left(\phi_{k,0} \!  -\!   {\theta_m}/{2}\right)\! .
\end{align}

As a sanity check, we can verify that the proposed model using first-order Taylor approximation is equivalent to the far-field channel model for uniform curved arrays. As shown in Fig. \ref{figure11} (b), the transmission distance difference under an impinging planar wave is given by the length of the segment AB. From elementary geometry,  we know
$ OB = 2 R \sin\left(\frac{\theta_m}{2}\right)  $,
$ 	\angle{AOC} = \frac{\pi}{2} - \phi_{k,0} $,
$ 	\angle{BOC} = \frac{\pi}{2} -   \frac{\theta_m}{2} $ ,
$ 	\angle{AOB} =\angle{BOC} - \angle{AOC} = \phi_{k,0} -     \frac{\theta_m}{2}  $,
and therefore obtain
\begin{align}
\begin{aligned}
	AB \!  \! &=\! OB \sin(\angle{AOB}) \!= \!2 R  \sin \!  \left( {\theta_m}/{2}  \right)   \sin\left(     \phi_{k,0} \!  -  \!    {\theta_m}/{2}    \right)\! \\
	& =  r_{k,0} - r_{k,0,\rm far}^{(m)}, 
\end{aligned}
\end{align}
which shows the consistency with our derivation.  

Now, based on (\ref{fist_order_taylor_UCA}) and (\ref{second_order_taylor_UCA}), we can  derive the theoretical Rayleigh distance in the considered uniform curved array systems and investigate the impact of the curvature angle $\theta$. The maximal deviation of the phase caused by the far-field approximation across the whole array is given by\cite{zhi2024performance}
\begin{align}
\begin{aligned}
&	{\rm Diff}= \max_m  \left[          
{2\pi}{} \left(r_{k,0, \rm near}^{(m)} -  r_{k,0, \rm far}^{(m) }  \right)      /\lambda
\right] \\
&=\!  \max_m \!  \left[          
{4\pi}{}   {R^2}{} \sin ^2\left({\theta_m}/{2}\right) \cos ^2\left(\phi_{k,0}-{\theta_m}/{2}\right) /( \lambda  r_{k,0} )
\right]\\
& = \max_m  \left[          
\frac{ 2\pi}{ \lambda}   \frac{R^2}{2r_{k,0}}      \left(\sin \phi_{k,0}+\sin \left(\theta_m-\phi_{k,0}\right) \right)^2
\right] .
\end{aligned}
\end{align}
By  letting $	{\rm Diff}={\pi}/{8}$, we obtain the theoretical Rayleigh distance  
\begin{align}
	r_{\rm ray} \!=\!  \max_m  \!    \left[          
	\frac{8}{\lambda}    \frac{L^2}{\theta^2}    \left(\sin \phi_{k,0} \!+\!    \sin \left(   \frac{m \theta}{M-1}         -\phi_{k,0}\right) \right)^2
	\right] 
\end{align}
where $m=-\frac{M-1}{2}, ...., \frac{M-1}{2}$.  Meanwhile, we can numerically calculate the area of near-field region as
\begin{align}
	\mathrm{S}_{\rm near}=\frac{1}{2} \int_{-\pi / 2}^{\pi / 2} r_{\text {ray }}(\phi_{k,0})^2 d \phi_{k,0}.
\end{align}

\begin{proposition}\label{corollary1}
	For a uniform  curved array with fixed arc length $L=(M-1)d$, as array curvature angle $\theta$ increases, the Rayleigh distance  for normal incident direction $\phi_{k,0}=0$ reduces while the Rayleigh distance for side incident direction $\phi_{k,0}=\pi/2$ increases. 
\end{proposition}

\itshape Proof: \upshape
	Note that $ R = \frac{L}{\theta} $ and $-\frac{\theta}{2}\leq \frac{m\theta}{M-1}  \leq \frac{\theta}{2}$. When $\phi_{k,0}=0$, we have
	\begin{align}
	r_{\rm ray}^{\rm normal}=          {2}    \left[  2R \sin({\theta}/{2})  \right]^2 /{\lambda}  \triangleq    {2 (\tilde{D}^{(\phi_{k,0}=0)} )^2   }/{\lambda}  ,
	\end{align}
which is proportional to the projected array aperture $  \tilde{D}^{(\phi_{k,0}=0)} $ observed in front of the array (in  $\phi_{k,0}=0$). Since $0\leq \theta  \leq \pi$,  $ r_{\rm ray}^{\rm normal} $ is a decreasing function of $\theta$. By contrast, when $\phi_{k,0}=\pi/2$, we have
	\begin{align}\label{side_rayleigh_UCA}
	r_{\rm ray}^{\rm side}=         \frac{8}{\lambda}     (  R   (  1- \cos( {\theta}/{2} )    )   )^2    \triangleq   {8(\tilde{D}^{(\phi_{k,0}=\pi/2)} )^2   }/{\lambda} ,
\end{align}
which is proportional to the projected array aperture  $ \tilde{D}^{(\phi_{k,0}=\pi/2)} $ observed in the side of the array (in   $\phi_{k,0}=\pi/2$). It is an increasing function of $\theta$.
$ \hfill \blacksquare $

Proposition \ref{corollary1} shows that the considered model effectively characterizes the impact of the array curvature  on the near-field region. As the array bends, users from the cell side will benefit from stronger near-field effects while the users in the cell center will suffer. 
Since near-field propagation can be exploited to mitigate multiuser interference even for users in LoS with the same AoAs, a curved array can be beneficial to improve the communication performance for users at the edge of the cell and enhance the coverage. It is also worth noting that when $\theta\to 0$, the derived Rayleigh distance will degrade to the conventional Rayleigh distance for ULA.

To demonstrate the effectiveness of the obtained insights, we also investigate the impact of array curvature  on the effective Rayleigh distance \cite{wu2023enabling}, defined in terms of the beamforming gain between near and far-field channels. We focus on the following normalized near-field steering vector with non-uniform phases:
\begin{align}
		\mathbf{a}_{\text {near}}(r_{k,0}, \phi_{k,0})=\left[e^{-j \kappa r_{k,0}^{(1)}   }, \ldots, e^{-j \kappa  r_{k,0}^{(M)}}     \right]^T /\sqrt{M} ,
\end{align}
where $\kappa\triangleq \frac{2\pi}{\lambda}$ is the wavenumber. Then, the near-field beamfocusing gain in the distance domain is given by
\begin{align}
		& g\left(r_1, r_2, \phi_{k,0}\right)=\left|\mathbf{a}_{\text {near}}^H\left(r_1, \phi_{k,0}\right) \mathbf{a}_{\text {near}}\left(r_2, \phi_{k,0}\right)\right|   \nonumber\\
		&=\frac{1}{M}\left|\sum\nolimits_{m=1}^M \exp \left\{j \kappa r_1^{(m)}-j \kappa r_2^{(m)}\right\}\right|    \nonumber\\
		& \overset{(a)}{\approx} \!\!     \frac{1}{M}   \!   \!     \left|\sum_{m=1}^M \!  \exp \!     \left\{   \!  j \kappa  \!   \left(\frac{R^2}{2 r_1} \! -    \frac{R^2}{2 r_2}\right)   \! \left[   \sin \phi_{k,0}  \!+\!  \sin \left(\theta_m  \!  -\!    \phi_{k,0}\right)    \right] ^2    \!        \right\}\right|       \nonumber\\
		& \overset{(b)}{=} \frac{1}{M}  \bigg|\sum\nolimits_{m=1}^M \exp \bigg\{j    \beta_{R, r_1, r_2}       \Big[     -\frac{1}{2}    {\cos \left(  2  \left(  \theta_m-\phi_{k,0}  \right)   \right)}        \nonumber\\
		&  +2 \sin \phi_{k,0} \cos \left(  \frac{\pi}{2} -  \theta_m  + \phi_{k,0}        \right)\Big]      \bigg\}\bigg|,
\end{align}
where $ (a)  $  applies (\ref{second_order_taylor_UCA}) and $(b)$ introduces $ \beta_{R, r_1, r_2} \triangleq \kappa (\frac{R^2}{2 r_1}-\frac{R^2}{2 r_2}  ) $ and applies some trigonometric identities.

%

\begin{figure}[t]
	\centering
	\includegraphics[width= 0.35\textwidth]{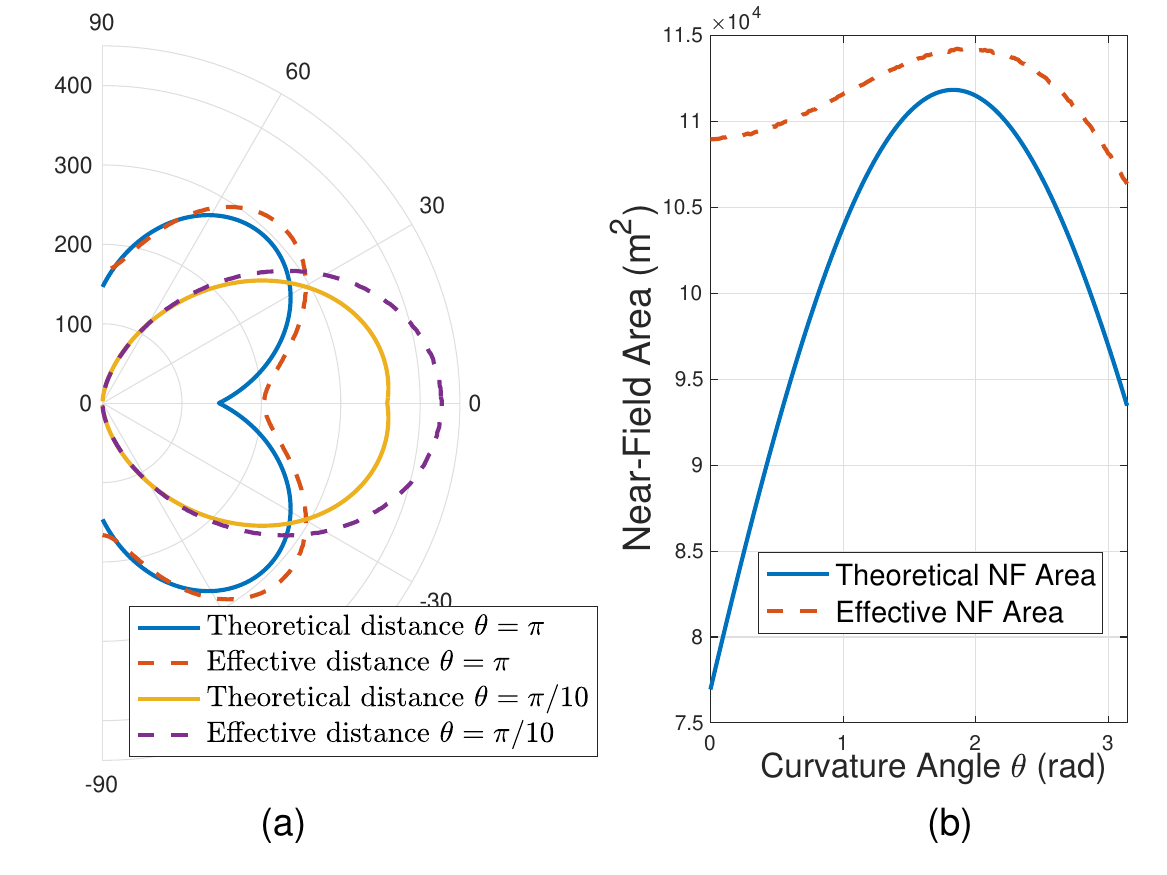}
			\vspace{-15pt}
	\caption{The impact of array curvature on (a) theoretical and effective Rayleigh distances; (b) near-field  area.}
			\vspace{-15pt}
	\label{figure18}
\end{figure}

Based on the     Jacobi-Anger expansion of Bessel functions where $ e^{j \beta \cos \gamma} = \sum_{n=-\infty}^{n=\infty} j^n J_n(\beta) e^{j n \gamma} $, we obtain (\ref{eff_rayleigh_dis}) on the bottom of the next page. When $\phi_{k,0}\to 0$, we have the approximation $ g\left(r_1, r_2, \phi_{k,0}\right)\approx  \left|  J_{0}\left(\frac{1}{2} \beta_{R_1 r_1 r_2}\right)  \right| $ by neglecting higher-order Bessel terms under large $M$, which is the same as  \cite[Eq. (42)]{wu2023enabling}. For larger incident angles, this  approximation does not hold in general. These observations unveil that the curved array has more complicated  effective Rayleigh distances which are functions of the incident angle $\phi_{k,0}$, showing the non-trivial impact of the curvature parameter $\theta$ on the near-field region for different directions.
\begin{figure*}[b]
	\hrule
\begin{align}\label{eff_rayleigh_dis}
		&g\left(r_1, r_2, \phi_{k,0}\right)    \nonumber   \\
		& =\frac{1}{M}\left|\sum_{m=1}^M \sum_{n_1=-\infty}^{n_1=\infty}\left\{j^{n_1} J_{n_1}\left(-\frac{1}{2} \beta_{R_1 r_1 r_2}\right) e^{-j n_1 2 \phi_{k,0}}\right\} e^{j n_1 2 \theta_m} \sum_{n_2=-\infty}^{n_2=\infty}\left\{j^{n_2} J_{n_2}\left(2 \sin \phi_{k,0} \beta_{R_1 r_1 r_2}\right) e^{j n_2\left(\frac{\pi}{2}+\phi_{k,0}\right)}\right\} e^{-j n_2 \theta_m}\right| \nonumber \\
		& =\left|\sum_{n_1=-\infty}^{n_1=\infty} \sum_{n_2=-\infty}^{n_2=\infty}\left\{j^{n_1+n_2} J_{n_1}\left(-\frac{1}{2} \beta_{R_1 r_1 r_2}\right) J_{n_2}\left(2 \sin \phi_{k,0} \beta_{R_1 r_1 r_2}\right) e^{j n_2\left(\frac{\pi}{2}+\phi_{k,0}\right)-j n_1 2 \phi_{k,0}}\right\} \frac{1}{M} \sum_{m=1}^M e^{j\left\{2 n_1-n_2\right\} \theta_m}\right|.
\end{align}
\end{figure*}

Based on $ g\left(r_1, r_2, \phi_{k,0}\right) $, we can define the effective Rayleigh distance as follows
\begin{align}\label{beamMis}
	{r}_{\rm ray}^{\text{eff}} =  \arg\min_{r_1}   g\left(r_1, \infty, \phi_{k,0}\right) > 0.95.
\end{align}
This means that, if one uses the far-field steering vector as a beamformer for users within this distance, the loss with respect to the optimal beamforming gain is more than $5\%$.

{\itshape{Numerical Example:}\upshape}  Now, we showcase some numerical results to validate the unveiled theoretical insights with $M=128$ at $6.8$ GHz.   It can be seen from Fig.   \ref{figure18} (a) that, as the curvature angle increases, the Rayleigh distance in the normal direction shrinks, but the Rayleigh distance in side direction increases, consistent with Proposition \ref{corollary1}. Besides, we observe that the derived theoretical Rayleigh distance based on the error in Taylor approximations has good consistency with the effective Rayleigh distance defined in terms of the beamforming loss \ref{beamMis}. 

In Fig. \ref{figure18} (b), we show the area of the near field region based on the Rayleigh distance. It can be seen that there exists an optimal curvature angle that maximizes the area. This is because the change of near-field region has a different tendency for different directions and thus the overall area is not a monotone function of $\theta$. This observation agrees with the results in Fig. \ref{figure18} (a), showing a performance trade-off between different directions for curved arrays.

\subsection{Uniform Cylindrical Array}\label{uniformCylindricalArray}

\begin{figure}[t]
	\centering
	\includegraphics[width= 0.35\textwidth]{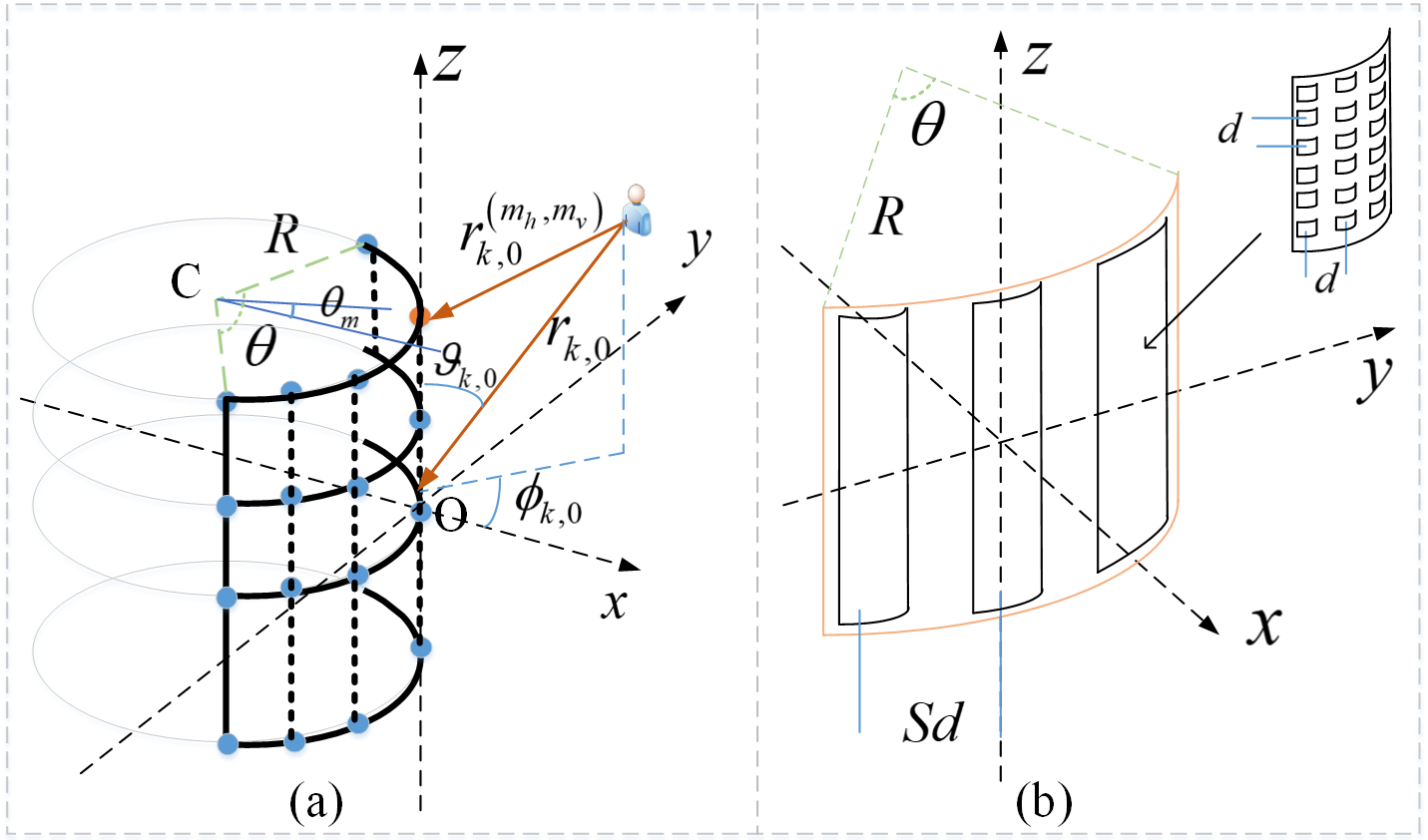}
	\DeclareGraphicsExtensions.
				\vspace{-15pt}
	\caption{The 3D curved array: (a) uniform cylindrical array; (b) modular cylindrical array.}
				\vspace{-15pt}
	\label{figure13}
\end{figure}

\begin{figure}[t]
	\centering
	\begin{minipage}{0.3\linewidth}
		\centering
		\includegraphics[width=\linewidth]{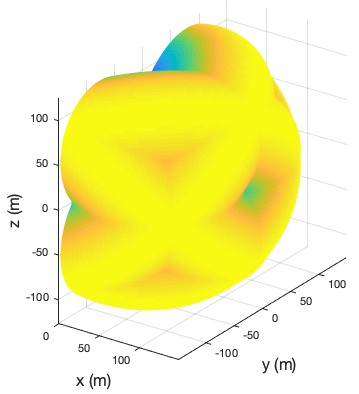}
		\centerline{\footnotesize (a)}
	\end{minipage}
\;\;\;
	\begin{minipage}{0.3\linewidth}
		\centering
		\includegraphics[width=\linewidth]{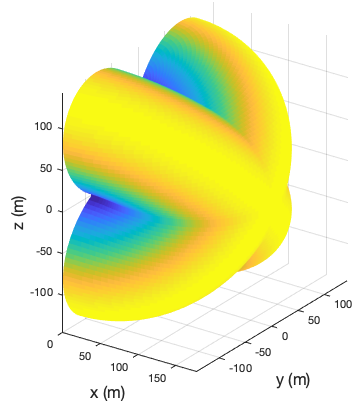}
		\centerline{\footnotesize (b) }
	\end{minipage}
	\caption{Near-field boundaries in the 3D space: (a)  uniform cylindrical array with  $\theta=\pi$; (b) conventional UPA.}
				\vspace{-15pt}
	\label{figure19}
\end{figure}

Next, we move to the three-dimensional (3D) space with uniform cylindrical arrays, as shown in Fig. \ref{figure13} (a). The cylindrical array can be viewed as an aggregation of $ M_v $ 2D curved arrays that are uniformly positioned along the z-axis with spacing $d$. Each horizontal curved array is a uniform curved array with $ M_h $ antennas. Thus, we have $M=M_hM_v$. Let $-\frac{M_h-1}{2} \leq m_h  \leq \frac{M_h-1}{2}$ and $-\frac{M_v-1}{2} \leq  m_v \leq  \frac{M_v-1}{2} $ denote the antenna indices in horizontal and vertical directions, respectively. The  curvature angle of the $m_h$-th horizontal antenna can be obtained as $\theta_{m_h} =m_h \frac{\theta}{M_h-1}$.   Setting the center of the array as origin, the 3D locations of the $({m_h},m_v)$-th antenna and   user $k$ can be  expressed, respectively,  as follows 
\begin{align} 
		&\mathbf{p}_{m_h,m_v} = \left[R\left(\cos \theta_{m_h} -1\right), R \sin \theta_{m_h}, m_v d\right],\\
		&\mathbf{u}_k =  [r_{k,0} \sin \vartheta_{k,0} \cos \phi_{k,0}, r_{k,0} \sin \vartheta_{k,0} \sin \phi_{k,0},    r_{k,0} \cos \vartheta_{k,0}   ] ,  \nonumber
\end{align} 
where $\vartheta_{k,0} $ denotes the vertical AoA from the center of the user to the center of the array. 
Then, the distance between  user $k$ and $({m_h},m_v)$-th antenna   is given by
\begin{align}\label{cylindrical_distance}
	&	r_{k,0}^{({m_h}, m_v)} = \left\|\mathbf{u}_k-\mathbf{p}_{{m_h},m_v}\right\|  \nonumber\\
	& \overset{(c)}{\approx} \!  r_{k,0} \!-  \!  R \sin \vartheta_{k,0}\left[\cos \left(\theta_{m_h} \!-\!   \phi_{k,0}\right)   \!-\!  \cos \phi_{k,0}\right]  \!-\!  m_v d \cos \vartheta_{k,0} \nonumber\\
	& +R^2 \frac{2\left(1 \!-\!    \cos \theta_{m_h}\right)  \!-\!     \sin ^2 \vartheta_{k,0}\left[  \cos \left(\theta_{m_h}  \!-\!        \phi_{k,0}\right)  \!-\!     \cos \phi_{k,0}  \right]^2}{2 r_{k,0}} \nonumber\\
	& -\frac{R}{r_{k,0}} m_v d \sin \vartheta_{k,0} \cos \vartheta_{k,0}\left[   \cos \left(\theta_{m_h}-\phi_{k,0}\right)-\cos \phi_{k,0}\right]  \nonumber\\
	&+ {m_v^2 d^2 \sin ^2 \vartheta_{k,0}}/{2 r_{k,0}},
\end{align}
applying   second-order Taylor approximation for $(c)$.


Similar to the analysis of 2D curved arrays, we can again use (\ref{cylindrical_distance}) to calculate the maximal phase errors  across the   cylindrical array  and let it equal to $\frac{\pi}{8}$, deriving the theoretical Rayleigh distance as follows
	\begin{align}\label{3D_rayleigh_distance}
	&	r_{\rm ray} = \max _m  \frac{1}{\lambda}  \bigg\{ 
	8 R^2   \big\{   2\left(  1  -    \cos \theta_{m_h}\right)  \nonumber\\
	&\!-\!   \sin ^2 \vartheta_{k,0}  \!  \left[\cos \left(\theta_{m_h}   \!- \!  \phi_{k,0}\right)  \!-\!   \cos \phi_{k,0}\right]^2 \!\!\big\} \nonumber\\
	&+8(M_v \!-\!  1) R d\left|\sin \vartheta_{k,0} \cos \vartheta_{k,0}\left[\cos \left(\theta_{m_h} \!-\!   \phi_{k,0}\right)-\cos \phi_{k,0}\right]  \right|    \nonumber\\
	&+2(M_v-1)^2 d^2 \sin ^2 \vartheta_{k,0}
	\bigg\}.
\end{align}

Based on (\ref{3D_rayleigh_distance}), we  obtain the following proposition.
\begin{proposition}\label{corollary2}
	For a uniform cylindrical array with fixed horizontal arc length $L_h$ and fixed vertical length $(M_v-1)d$, as array curvature angle $\theta$ increases, the Rayleigh distance  for normal incident direction ($\vartheta_{k,0}=\pi / 2$ and $\phi_{k,0}=0$)  reduces while the Rayleigh distance for side incident direction ($\vartheta_{k,0}=\pi / 2$ and $\phi_{k,0}=\pi/2$) increases. 
\end{proposition}

\itshape Proof: \upshape
 When the signal impinges perpendicularly to the center of the array, substituting (\ref{3D_rayleigh_distance}) with $\vartheta_{k,0}=\pi / 2$ and $\phi_{k,0}=0$, we have
 \begin{align}
 	 r_{\rm ray} ^{\rm normal}\!=\!  \frac{2}{\lambda} \! \left[   \left(   2 R \sin \left( {\theta}/{2}\right)  \right)  ^2+((M_v-1) d)^2    \right] \!=\frac{2\tilde{D}_{\rm n}^2}{\lambda},
 \end{align}
which is proportional to the effective aperture $ \tilde{D}_{\rm n} $ of the 3D array observed from the normal direction of $\vartheta_{k,0}=\pi / 2$ and $\phi_{k,0}=0$. This effective aperture decreases as $\theta$ increases.
 
 For side direction,  substituting (\ref{3D_rayleigh_distance}) with $\vartheta_{k,0}=\pi / 2$ and $\phi_{k,0}=\pi/2$, we have
  \begin{align}
 		 r_{\rm ray}^{\rm side} {=}    \frac{2}{\lambda}   \left[    \left(   2 R \left(1-\cos({\theta}/{2})\right) \right)  ^2+((M_v-1) d)^2\right]  ,
 \end{align}
 whose tendency with $\theta$ is the same as (\ref{side_rayleigh_UCA}).
$ \hfill \blacksquare $

\itshape Numerical Example: \upshape Again, we use some numerical results to illustrate the non-trivial impact of 3D array curvature angle on near-field region with $M_h=M_v=64$ at $6.8$ GHz.
It can be seen from Fig. \ref{figure19} that the cylindrical array expands the near-field region in the side direction while reduces the near-field effect in the normal direction. There also exists an optimal curvature angle of the cylindrical array in terms of the volume of the near-field region, which is not shown here due to the limit of page.

\subsection{Modular Cylindrical Array}

To mitigate the negative impact of curved array on the normal direction and further enlarge the near-field region, we propose a general array geometry called modular cylindrical array as illustrated in Fig. \ref{figure13} (b).  Assume that there are $I$ horizontal  tiles with $M_i=\frac{M_h}{I}  M_v$ antennas in each tile. As a sparse array,  denote the antenna spacing in each tile  and the spacing between horizontal subarrays as   $d$ and $D = Sd$, respectively, where $S\geq M_h/I$.  When $S=M_h/I$, the horizontal tiles are closed spaced, i.e., becoming uniform cylindrical arrays in \ref{uniformCylindricalArray} .  

Then, the horizontal dimension of the array is given by $L_h = (I-1)Sd  + ( \frac{M_h}{I}-1  )   d$  and the radius of the array is  $ R = \frac{L_h}{\theta}  $.
Let $ i_h = -\frac{I-1}{2}, \cdots, \frac{I-1}{2} $, $ m_h = -\frac{M_h/I-1}{2}, \cdots, \frac{M_h/I-1}{2} $, and $ m_v= -\frac{M_v-1}{2}, \cdots,  \frac{M_v-1}{2} $ denote the indices of subarray, horizontal antenna, and vertical antenna, respectively. 
The curvature angle of the central location of the $(m_h,m_v)$-th antenna in $i_h$-th tile is given by 
\begin{align}
	\theta_{i_h,m_h} =    \frac{\theta}{   (I-1)S + \frac{M_h}{I} -1   }  \left( i_h  S +  {m_h}  \right)  . 
\end{align}
Accordingly, the location of the $(m_h,m_v)$-th antenna in the  $i_h$-th tile is given by
\begin{align}
		 \mathbf{p}_{i_h,m_h,m_v}  \!  = \! [    R\left(\cos \theta_{i_h,m_h}    -1\right),      R \sin \theta_{i_h,m_h}   ,      m_v  d      ]^T  .
\end{align}
Then, the distance between    user $k$ and  the $(m_h,m_v)$-th antenna in the $i_h$-th tile can be calculated as
\begin{align}
	\begin{aligned}
	&	r_{k,0}^{(i_h,m_h,m_v)} = \left\|  \mathbf{u}_k    -  \mathbf{p}_{i_h,m_h,m_v}     \right\| \\
		&= \Big\{   r_{k,0}^2 + 2 {R^2}  \left(1  -   \cos  \theta_{i_h,m_h}       \right) \\
		&-2 {R}{r_{k,0}} \sin \vartheta_{k,0}     \left[      \cos \left(   \theta_{i_h,m_h}     -  \phi_{k,0}    \right)   -    \cos \phi_{k,0}     \right] \\
		&-{2}{r_{k,0}}  m_v    d \cos \vartheta_{k,0}   +   {   (  m_v )^2 d^2}   \Big\}^{\frac{1}{2}}.
	\end{aligned}
\end{align}

\begin{proposition}\label{corollary_modular}
		For the modular cylindrical array, the Rayleigh distances in the normal incident direction, i.e., $\vartheta_{k,0}=\pi/2$ and $\phi_{k,0}=0$, and in the side incident direction, i.e., $\vartheta_{k,0}=\pi/2$ and $\phi_{k,0}=\pi/2$, are respectively given by
	\begin{align}\label{modular_rayleigh_normal}
 r_{\rm ray}^{\rm normal}
=
\frac{2}{\lambda}
\left\{
\left(
\frac{2  L_h}{\theta}
\sin \frac{\theta}{2}
\right)^2
+
\left[   (M_v-1)d   \right]^2
\right\},
	\end{align}
	and
	\begin{align}\label{modular_rayleigh_side}
r_{\rm ray}^{\rm side}
=
\frac{2}{\lambda}
\left\{
\left[
\frac{2L_h}{\theta}
\left(1-\cos\frac{\theta}{2}\right)
\right]^2
+
\left[     (M_v-1)d    \right]^2
\right\}.
	\end{align}
\end{proposition}

\itshape Proof: \upshape
It follows a similar proof as Proposition \ref{corollary2}.
$ \hfill \blacksquare $

It can be observed from Proposition \ref{corollary_modular} that an increase of the subarray spacing $S$ enlarges $L_h$ and hence expands the near-field region in both directions. Moreover, for fixed $L_h$ and vertical height, increasing the curvature angle $\theta$ reduces the Rayleigh distance in the normal direction but increases  it   in the side direction.

\begin{remark}
The parameters of $\theta$, $I$, $S$ for the proposed modular cylindrical array  can be designed to improve the performance of near-field communication, achieving a balance between extending near-field region and controlling the effect of grating lobes. This will be discussed in Section \ref{section5}.
\end{remark}

\section{General Channel Estimation Algorithms}\label{section4}
An accurate knowledge of the user channels is fundamental to materialize the multiplexing gain of multiuser MIMO\cite{davoodi2016aligned}.  
In this section, we focus on the estimation of  users' channels $\mathbf{h}_k$, $\forall k$ from uplink pilots received at the BS (e.g., SRS pilots as in 5G-NR\cite{3gpp2018nr}).

We denote the uplink analog RF combiner at the BS by   $\mathbf{V}_{\rm ul} \in \mathbb{C}^{ M_{RF} \times M}$ whose non-zero elements are independently randomly generated with constant modulus. Consistent with the current wireless standards\cite{3gpp2018nr}, orthogonal pilot signals are allocated to $ K $ users, where $  \mathbf{s}_k\in \mathbb{C}^{\tau\times 1} $ denotes the pilot sequence of user $k$, and $\mathbf{s}_k^H \mathbf{s}_j= \delta_{k,j}$. 
Let $[\mathbf{s}_k]_i = s_{k,i}$. At time slot $i$, the received pilot $\mathbf{y}_i \in\mathbb{C}^{M_{RF}\times 1}$ at the BS is
\begin{align}
	\mathbf{y}_{i} = \sum\nolimits_{k=1}^{K} \sqrt{\tau p_{\rm ul, k}} \;\mathbf{V}_{\rm ul} \mathbf{h}_k  s_{k,i} +  \mathbf{V}_{\rm ul} \mathbf{n}_i,
\end{align}
where $ \mathbf{n}_i   \sim \mathcal{C N}(\mathbf{0},\sigma^2 \mathbf{I}_M)$ is the Gaussian complex noise at the BS in time slot $i$ and $p_{\rm ul, k}$ is the energy per symbol of the pilot sequence of user $k$. Due to the sub-connected property $\mathbf{V}_{\mathrm{ul}} \mathbf{V}_{\mathrm{ul}}^H=\mathbf{I}_{M_{\mathrm{RF}}}$, we have $ \mathbf{V}_{\rm ul} \mathbf{n}_i  \sim \mathcal{C N}(\mathbf{0},\sigma^2 \mathbf{I}_{M_{RF}})$.

After $\tau $ slots, we can stack the received pilots as $\mathbf{Y} = [\mathbf{y}_1, \cdots, \mathbf{y}_\tau] \in\mathbb{C}^{M_{RF}\times \tau}$, which can be expressed as follows
\begin{align}
	\mathbf{Y}  = \mathbf{V}_{\rm ul}  \mathbf{H}_{\rm ul} \mathbf{P}_{\rm ul}   \mathbf{S} + \mathbf{V}_{\rm ul} \mathbf{N},
\end{align}
where $ \mathbf{H}_{\rm ul} \triangleq  [\mathbf{h}_1  , \cdots, \mathbf{h}_K ] \in\mathbb{C}^{M\times K}$, $\mathbf{P}_{\rm ul}  \triangleq \operatorname{diag}\{   \sqrt{\tau p_{\rm ul, 1}}, \cdots , \sqrt{\tau p_{\rm ul,K}}          \}$,  $\mathbf{S} \triangleq   [\mathbf{s}_1, \cdots, \mathbf{s}_K]^T   \in\mathbb{C}^{K\times \tau}    $, and $\mathbf{N}  \triangleq   [\mathbf{n}_1, \cdots, \mathbf{n}_\tau]  \in\mathbb{C}^{M\times \tau}$.
Using the pilot orthogonality, we obtain the observation $ \mathbf{y}_{k} \in \mathbb{C}^{M_{RF} \times 1 }$ of user $k$ as
\begin{align}\label{sdafsdf}
	\mathbf{y}_{k} =  \frac{  \mathbf{Y} \mathbf{s}_k^*    }{\sqrt{\tau p_{\rm ul, k}}}     &=   \mathbf{V}_{\rm ul} \mathbf{h}_k  +  \frac{\mathbf{V}_{\rm ul} \mathbf{N} \mathbf{s}_k^*   }{\sqrt{\tau p_{\rm ul, k}}}   \triangleq  \mathbf{V}_{\rm ul} \mathbf{h}_k  +  \tilde{\mathbf{n}} ,
\end{align}
where $\tilde{\mathbf{n}}    \sim \mathcal{C N}(\mathbf{0},  \sigma_{\rm bs}^2  \mathbf{I}) $, $ \sigma_{\rm bs}^2  \triangleq  \frac{\sigma^2  }{ {\tau p_{\rm ul, k}}}    $, is the equivalent noise.

The main challenge of estimating $\mathbf{h}_k$ based on (\ref{sdafsdf}) is that $\mathbf{V}_{\rm ul} $ is typically a very low-dimensional projection, because $M_{RF} \ll M$. Hence, (\ref{sdafsdf}) is an underdetermined linear observation that yields infinite possible solutions even in the absence of noise. This dimensionality bottleneck problem is typically addressed by invoking sparsity in the framework of compressed sensing, which requires a good choice of a suitable ``sparsifying dictionary’’. In the far-field case, DFT dictionaries typically work well.  In the near-field case, the choice of a generally good sparsifying dictionary is difficult, especially considering the variation of array geometries and the presence of spatial non-stationarity.

Therefore, without relying on a specific codebook, we propose to solve the following antenna-domain channel estimation problem
\begin{align}\label{Lproblem}
	\begin{aligned}
		& \hat{\mathbf{h}}_k  \!=\! 
		\arg \min _{     \mathbf{h}_k    }
		\left\|     {\mathbf{y}}_{k}   -
		\mathbf{V}_{\rm ul}  \mathbf{h}_k 
		\right\|_2^2 \! + \underbrace{\lambda  \left\|     {\mathbf{h}}_{k}   -
			\mathcal{R}_{\boldsymbol{\Theta}}\left\{\mathbf{h}_k \right\}
			\right\|_2^2}_{\text{Learned Regularizer}}
	\end{aligned}
\end{align}
where $  \mathcal{R}_{\boldsymbol{\Theta}}\left\{  \cdot  \right\} $  is a regularization function that can be learned using a neural network with parameter $\boldsymbol{\Theta}$.   This learned regularizer will help  the estimation algorithm to generate estimates that satisfy certain inherent features of the channel  $\mathbf{h}_k$. 
In particular, the channel features capture the physics of the propagation environment which  do not reduce to a simple sparsity assumptions modelled by explicit codebooks. 


\begin{figure}[t]
	\centering
	\includegraphics[width= 0.49\textwidth]{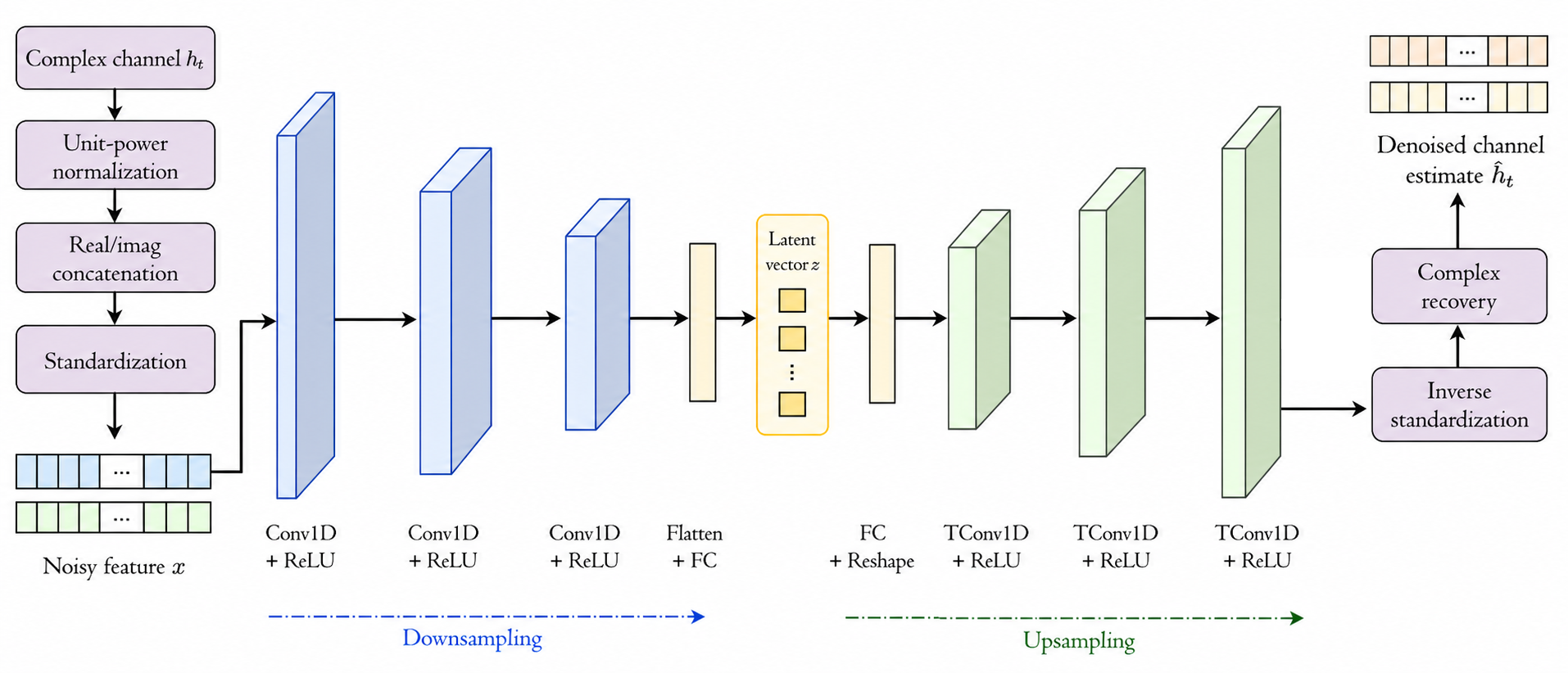}
	\DeclareGraphicsExtensions.
			\vspace{-25pt}
	\caption{Illustration of the considered  denosing AE network.}
	\label{figure41}
				\vspace{-15pt}
\end{figure}

\subsection{CNN-Based  Denoising Autoencoder}
As illustrated in Fig. \ref{figure41}, we design a convolutional neural network (CNN)-based denoising autoencoder  
that implicitly learns an efficient latent representation of the antenna-domain channel. Here ``efficient’’ means that the number of coefficients in the latent space is typically much smaller than the channel dimension, thus generalizing the concept of ``sparsity’’.  After training, the network has the ability to capture the inherent feature of the channel such that the second term in the estimation metric (\ref{Lproblem}) can be interpreted as a penalty term that forces the solution $ \hat{\mathbf{h}}_k$ to be close to the manifold of typical channels learned by the network.

\subsubsection{Preprocessing} Since the training is performed across the whole service area with distinct pathloss, before being fed into the network, each complex channel sample $\mathbf{h}_{(s)}\in\mathbb{C}^{M}$, $1\leq s \leq N_{\mathrm{train}}$, is normalized to $\bar{\mathbf{h}}_{(s)}$ that has  unit average power and then converted into a real-valued Cartesian representation as
\begin{align}
	\mathbf{x}_{(s)}
	=
	\left[
	\Re\{\bar{\mathbf{h}}_{{(s)}}\}^{T},
	\Im\{\bar{\mathbf{h}}_{{(s)}}\}^{T}
	\right]^{T}
	\in\mathbb{R}^{2M}.
\end{align}
The feature vector $\mathbf{x}$ is then standardized using the statistics computed from the training set. This preprocessing helps the network to learn the channel manifold across a large area.

\subsubsection{Encoder}  The encoder reshapes the input vector into a two-channel sequence of size $2\times M$, where the two channels correspond to the real and imaginary parts of the channel. It then applies three one-dimensional convolutional layers with a stride of two to progressively extract spatial features along the antenna dimension. The output of the last convolutional layer is flattened and mapped to a latent vector through a fully connected layer. Mathematically, the encoder with input feature ${\mathbf{x}}_{(s)}$ can be written as
\begin{align}
	\mathbf{z}_{(s)}
	=
	f_{\rm enc}
	\left(
	\mathbf{x}_{(s)};
	\boldsymbol{\Theta}_{\rm enc}
	\right),
\end{align}
where $ \boldsymbol{\Theta}_{\rm enc} $ includes the training parameters for encoder network and $\mathbf{z}_{(s)}$ is the latent representation of sample $s$. 

\subsubsection{Decoder}  The decoder mirrors the encoder structure. It first maps the latent vector back to the deepest feature dimension and then uses transposed convolutional layers to recover the original channel dimension.  The final output is transformed back into the complex domain, yielding the $s$-th reconstructed channel estimate
\begin{align}
	\hat{\mathbf{x}}_{(s)}
	=
	f_{\rm dec}
	\left(
	\mathbf{z}_{(s)};
	\boldsymbol{\Theta}_{\rm dec}
	\right),
\end{align}
where $f_{\rm dec}(\cdot)$ denotes the decoder and $\boldsymbol{\Theta}_{\rm dec}$ represents its trainable parameters. Finally, $\hat{\mathbf{x}}_{(s)}$ is inversely standardized and transformed back into the complex domain to obtain the reconstructed unit-power channel $	\hat{\bar{\mathbf{h}}}_{(s)} $.

\subsubsection{Loss Function} The overall network can be denoted by $  	\hat{\bar{\mathbf{h}}}_{(s)} = 	\mathcal{R}_{\boldsymbol{\Theta}}(\bar{\mathbf{h}}_{(s)} )  $  with trainable parameters
$\boldsymbol{\Theta} =  \{\boldsymbol{\Theta}_{\rm enc},\boldsymbol{\Theta}_{\rm dec}\}$.  The reconstruction loss is considered to preserve the channel manifold representation as follows
\begin{align}
	\mathcal{L}_{\rm recon}^{(s)}
	=
	\frac{1}{2}
	\left\|
	\mathcal{R}_{\boldsymbol{\Theta}}
	\left(\bar{\mathbf{h}}_{(s)}\right)
	-
	\bar{\mathbf{h}}_{(s)}
	\right\|_2^2 .
\end{align}

Besides, we consider the following denoising loss    so that the trained autoencoder can effectively denoise
\begin{align}
	\mathcal{L}_{\rm denoiser}^{(s)}
	=
	\frac{1}{2}
	\left\|
	\mathcal{R}_{\boldsymbol{\Theta}}
	\left( \bar{\mathbf{h}}_{(s)}  +  \breve{ \mathbf{n} } \right)
	-
\bar{	\mathbf{h}}_{(s)}
	\right\|_2^2 ,
\end{align}
where  $\breve{\mathbf{n}}\sim\mathcal{CN}(\mathbf{0},\sigma_{\rm train}^2\mathbf{I})$ denotes the artificially injected complex Gaussian noise. To be general, we do not train multiple network for different SNRs. Only one AE is trained with noise  variance randomly sampled by  $ \sigma_{\rm train}^2  = \frac{1}{  {\rm SNR}_{T}    }$ where 
$ 	{{\rm SNR}_{T}  }   \sim\mathcal{U} \left(      {{\rm SNR}_{\rm min}}  ,     {{\rm SNR}_{\rm max}}    \right) $.

The overall training objective is
\begin{align}
	 \boldsymbol{\Theta}^* =  \arg\min_{\boldsymbol{\Theta}}
	\sum_{s=1}^{N_{\mathrm{train}}} \frac{ \!  \alpha_{\rm ae} \mathcal{L}_{\rm recon}^{(s)}
		+
		(1 \!-\!   \alpha_{\rm ae})\mathcal{L}_{\rm denoiser}^{(s)}  }{N_{\mathrm{train}}},
\end{align}
where $\alpha_{\rm ae}$ balances the clean reconstruction ability and the denoising capability. 


\subsubsection{Training Dataset}
The proposed deep denoiser is trained offline using   channel samples generated from the considered   large-scale environment. For each training sample, the   locations of the user and scatterers are randomly drawn from a predefined three-dimensional spatial region. Given the sampled user and scatterer locations and the array geometry, the corresponding complex channel vector is generated according to the spherical-wave propagation model (\ref{channel_H}) with LoS and NLoS components.
For the LoS component, the distance-dependent phase shift and path attenuation between the user and each antenna element are explicitly calculated. For the NLoS components,  each path is calculated as a two-hop propagation link from the user to the scatterer and   from the scatterer to the array, with corresponding pathloss and phases. A random complex path coefficient is further introduced to account for the small-scale fading of each scattered path. In addition, non-stationary visibility masks are applied to both LoS and NLoS components to emulate the spatial non-stationarity of extremely large-scale arrays.

\subsection{Estimation Algorithms}
After offline training, the proposed network $ \mathcal{R}_{\boldsymbol{\Theta}^*}\left(\cdot\right) $   can be employed as a learned regularizer in the following channel estimation algorithms\footnote{Since the network is trained for unit-power channels, in the application, the input is normalized and the amplitude of the output is recovered by a least-squares (LS)-based scaling.}. 

\subsubsection{Gradient Descent}
To solve problem (\ref{Lproblem}), the simplistic method is gradient descent. The gradient of the objective function can be expressed as
\begin{align}\label{fsdgsdfhs}
	\begin{aligned}
		\nabla_{\mathbf{h}_k^*} f\left(\mathbf{h}_k\right)&= \mathbf{V}_{\rm ul} ^H\left(\mathbf{V}_{\rm ul}  \mathbf{h}_k-\mathbf{y}_k\right)\\
		&+ \lambda\left(\mathbf{I}-\mathbf{J}_{\mathcal{R}}\left(\mathbf{h}_k\right)\right)^H\left(\mathbf{h}_k-\mathcal{R}_{\boldsymbol{\Theta}^*}\left(\mathbf{h}_k\right)\right),
	\end{aligned}
\end{align}
where
$ 	\mathbf{J}_{\mathcal{R}}(\mathbf{h}_k)=\frac{\partial \mathcal{R}_{\boldsymbol{\Theta}^*}(\mathbf{h}_k)}{\partial \mathbf{h}_k^T} $
is the Jacobian matrix of the network and can be calculated based on backpropagation of the trained network. Then, the channel variable can be estimated iteratively as
\begin{align}\label{gradient_descent}
	\mathbf{h}_k^{(t+1)}=\mathbf{h}_k^{(t)}-{\xi} \nabla_{\mathbf{h}_k^*} f\left(\mathbf{h}_k^{(t)}\right) 
\end{align}
with step length ${\xi}$. While simplistic, as shown in (\ref{fsdgsdfhs}), it is worth noting that the method of gradient descent is sensitive to the choice of weight factor $\lambda$ and to the choice of the initial guess which may end up in a local minimum. A possible method is to build a grid of optional values for $\lambda$ and choose the optimal one. However, it will increase the complexity.

\subsubsection{AE-Aided PnP}
It can be seen that the gradient descent method requires the calculation of the gradient of the network. To reduce the complexity and accelerate the convergence,  we decouple the data term and the prior term into two separate subproblems, allowing closed-form solutions in each iteration. Specifically, at the $t$-th iteration, we first update an intermediate estimate by solving a quadratic
data-fidelity subproblem
\begin{align}
	\tilde{\mathbf{h}}_k^{(t+1)}
	&= \arg\min_{\mathbf{h}}
	\big\|\mathbf{y}_k - \mathbf{V}_{\rm ul} \mathbf{h}\big\|_2^2
	+ \mu \big\|\mathbf{h} - \mathbf{h}_k^{(t)}\big\|_2^2 \\
	&= \big(\mathbf{V}_{\rm ul} ^H \mathbf{V}_{\rm ul}  + \mu \mathbf{I}\big)^{-1}
	\big(\mathbf{V}_{\rm ul} ^H \mathbf{y}_k + \mu \mathbf{h}_k^{(t)}\big),
\end{align}
and then  apply the learned network  as a projection/denoiser:
$	\mathbf{h}_k^{(t+1)}	= \mathcal{R}_{\boldsymbol{\Theta}^*}\big(\tilde{\mathbf{h}}_k^{(t+1)}\big).$
This two-step scheme is in line with plug-and-play methods based on
half-quadratic splitting\cite{zhang2021plug}, where the network
$\mathcal{R}_{\boldsymbol{\Theta}^*}$ plays the role of an implicit prior via a learned
denoiser. Nevertheless, this method also suffers from the effective choice of weight factor  $\mu$.

\subsubsection{AE-Aided AMP}
To avoid the  manual tuning of weight factors, we apply an AE-based AMP method in which the balance between data  and prior terms  is adaptively adjusted by the residual recursion and the Onsager correction term. The algorithm is iterated in the following way
\begin{align}
	\mathbf{v}_k^{t} 
	&= \mathbf{h}_k^{t} + \mathbf{V}_{\rm ul} ^{H}\mathbf{r}_k^{t},
	\label{eq:aeamp_vk}\\
	\tilde{\mathbf{h}}_k^{t+1} 
	&= \mathcal{R}_{\boldsymbol{\Theta}^*}\big(\mathbf{v}_k^{t}\big),
	\label{eq:aeamp_denoise}\\
	\mathbf{r}_k^{t+1} 
	&= \mathbf{y}_k - \mathbf{V}_{\rm ul} \,\tilde{\mathbf{h}}_k^{t+1}
	+ \frac{\operatorname{div}\mathcal{R}_{\boldsymbol{\Theta}^*}(\mathbf{v}_k^{t})}
	{\delta}\,\mathbf{r}_k^{t},
	\label{eq:aeamp_residual}
\end{align}
where $\delta \triangleq M_{RF}/M$ is the measurement rate and $\mathbf{r}_k^t$ denotes the residual vector at the $t$-th iteration whose  initial value can be set as $     \mathbf{r}_k^{0} = \mathbf{y}_k $

Following the denoising-AMP framework\cite{metzler2016denoising}, the Onsager correction term requires the
divergence of the denoiser, which we approximate via a Monte-Carlo
estimator:
\begin{align}
	\begin{aligned}
		&	\operatorname{div}\mathcal{R}_{\boldsymbol{\Theta}^*}\big(\mathbf{v}_k^{t}\big)
		\approx  \frac{1}{K_{\mathrm{mc}}\varepsilon M}\\
		&\times 
		\sum\nolimits_{\ell=1}^{K_{\mathrm{mc}}}
		\big( \mathbf{u}^{(\ell)} \big)^{\!H}
		\Big(
		\mathcal{R}_{\boldsymbol{\Theta}^*}\big(\mathbf{v}_k^{t} 
		+ \varepsilon \mathbf{u}^{(\ell)}\big)
		- \mathcal{R}_{\boldsymbol{\Theta}^*}\big(\mathbf{v}_k^{t}\big)
		\Big),
	\end{aligned}
\end{align}
where $\mathbf{u}^{(\ell)}\sim \mathcal{CN}(\mathbf{0},\mathbf{I}_M)$ and
$\varepsilon>0$ is a small perturbation. To improve stability, we also adopt the following damping step
\begin{align}
	\mathbf{h}_k^{t+1}
	= (1-\gamma)\,\mathbf{h}_k^{t}
	+ \gamma\,\tilde{\mathbf{h}}_k^{t+1},
	\qquad \gamma\in(0,1].
	\label{eq:aeamp_damping}
\end{align}

{\itshape{State Evolution and Optimal Bound:}\upshape} 
To evaluate the performance and show its reasonability, we establish a lower bound for our scheme by assuming the ideal case where each element of matrix $\mathbf{V}_{\rm ul} $ is i.i.d. Gaussian with $ \mathcal{CN}( {0},  {1/M_{RF}})$. In this case, the AE-aided AMP admits the state evolution recursion
\begin{equation}
	\nu_{t+1}^2 \;=\; \sigma_{\rm amp}^2\;+\; \frac{1}{\delta}\,
	\mathrm{MSE}_t(\nu_t^2)
	\label{eq:se}
\end{equation}
initialized with $\nu_0^2 = \sigma_{\rm amp}^2+ \delta^{-1}M^{-1}\mathbb{E}\|\mathbf{h}_k\|_2^2$, where 
the per-component MSE is defined as
\begin{equation}
	\mathrm{MSE}_t(\nu_t^2) \;\triangleq\;
	\frac{1}{M}\,\mathbb{E}\!\left[\,
	\big\|\mathcal{R}_{\Theta^\star}(\mathbf{h}_k+\nu_t\mathbf{w}_t)
	- \mathbf{h}_k\big\|_2^2\right], 
	\label{eq:mse-def}
\end{equation}
with $\mathbf{w}_t \sim \mathcal{CN}(\mathbf{0}, \mathbf{I})$ and the expectation taken over both the CSI $\mathbf{h}_k$ and the noise $\mathbf{w}_t$. 
For fair, we set $\sigma_{\rm amp}^2  \triangleq   \sigma_{\rm bs}^2 / \delta$ since $\mathbf{V}_{\rm ul} \mathbf{V}_{\rm ul} ^H=\mathbf{I}$ under the considered model while $\mathbb{E}\{\mathbf{V}_{\rm ul} \mathbf{V}_{\rm ul} ^H\}=\mathbf{I}/\delta$  under the assumption of  $[ \mathbf{V}_{\rm ul} ]_{m_1,m_2} \sim  \mathcal{CN}( {0},  {1/M_{RF}})$.

The asymptotic estimation error of the AE-aided AMP is then characterized by the fixed point $\mathrm{MSE}_{\infty}(\nu_{\infty}^2)$. By replacing $\mathcal{R}_{\Theta^\star}$ with the ideal posterior-mean (MMSE) denoiser under the known channel prior,
\begin{equation}\label{posterieMean}
	\mathcal{R}_{\mathrm{mmse}}(\mathbf{v};\nu^2)
	\;=\; \mathbb{E}\!\left[\mathbf{h}_k\,\big|\,
	\mathbf{h}_k+\nu\mathbf{w}=\mathbf{v}\right],
\end{equation}
the SE recursion \eqref{eq:se} reduces, under the replica-symmetric ansatz\cite{guo2005randomly,metzler2016denoising}, to the self-consistent fixed-point equation
\begin{equation}
	\nu^2 \;=\; \sigma_{\rm amp}^2\;+\; \frac{1}{\delta}\,
	\operatorname{mmse}(\nu^2),
	\label{eq:replica-fp}
\end{equation}
where 
\begin{equation}\label{gshsrgr}
	\operatorname{mmse}(\nu^2)
	\;\triangleq\;
	\frac{1}{M}\,\mathbb{E}\!\left[\,
	\big\|\mathbb{E}[\mathbf{h}_k\mid\mathbf{h}_k+\nu\mathbf{w}]
	- \mathbf{h}_k\big\|_2^2\right].
\end{equation}

To calculate (\ref{gshsrgr}), we independently generate $Q$ large-scale channel states $\{\boldsymbol{\zeta}_q\}_{q=1}^{Q}$ from the considered environment, including the user location, scatterer locations, number of paths, and VR patterns, where $p(\boldsymbol{\zeta}_q)=1/Q$. Conditioned on a realization $\boldsymbol{\zeta}_q$,   channel  $\mathbf{h}_k$ is Gaussian distributed with 
$	\mathbf h_k\mid \boldsymbol{\zeta}_q \sim	\mathcal{CN}	\left(	\boldsymbol{\mu}_q, 	\boldsymbol{\Sigma}_q	\right)$.  Then, the channel is of mixed Gaussian distribution  
\begin{align}
p(\mathbf{h}_k) \approx  \frac{1}{Q} \sum\nolimits_{q=1}^Q \mathcal{C} \mathcal{N}\left(\mathbf{h}_k ; \boldsymbol{\mu}_q, \boldsymbol{\Sigma}_q\right).
\end{align}
Since $\mathbf{v}=\mathbf{h}_k+\nu\mathbf{w}$, we have $\mathbf{v} \mid \boldsymbol{\zeta}_q \sim  \mathcal{C N}\left(\boldsymbol{\mu}_q,      \mathbf{\Sigma}_q+\nu^2 \mathbf{I}      \right)$. Given $\boldsymbol{\zeta}_q $ and $\mathbf{v}$, we have the conditional posterior mean
\begin{align}
 \mathbb{E}[\mathbf{h}_k \mid \mathbf{v}, \boldsymbol{\zeta}_q] =  \boldsymbol{\mu}_q+\boldsymbol{\Sigma}_q\left(\boldsymbol{\Sigma}_q+\nu^2 \mathbf{I}\right)^{-1}\left(\mathbf{v}-\boldsymbol{\mu}_q\right).
\end{align}
The posterior mean (\ref{posterieMean}) therefore can be calculated as
\begin{align}\label{mmse_MC}
\mathbb{E}[\mathbf{h}_k \mid \mathbf{v}]=\sum\nolimits_{q=1}^Q p(  \boldsymbol{\zeta}_q    \mid \mathbf{v}) \mathbb{E}[\mathbf{h}_k \mid \mathbf{v}, \boldsymbol{\zeta}_q] ,
\end{align}
where 
\begin{align}
	\begin{aligned}
		&p( \boldsymbol{\zeta}_q  \mid \mathbf{v})=\frac{  p( \mathbf{v}   \mid \boldsymbol{\zeta}_q)     p(    \boldsymbol{\zeta}_q     )     }{    p(\mathbf{v})   }
		=\frac{  p( \mathbf{v}   \mid \boldsymbol{\zeta}_q)     p(\boldsymbol{\zeta}_q)     }{  \sum_{j=1}^{Q}     p(\mathbf{v},    \boldsymbol{\zeta}_j      )    }
		\\
		&=\frac{  p( \mathbf{v}   \mid \boldsymbol{\zeta}_q)     p(\boldsymbol{\zeta}_q)     }{  \sum_{j=1}^{Q}          p(\mathbf{v}   \!  \!     \mid   \boldsymbol{\zeta}_j      )   p(   \boldsymbol{\zeta}_j      )   }
		\!=\!   \frac{      \mathcal{C N}      \left(   \mathbf{v}  ;    \boldsymbol{\mu}_q, \boldsymbol{\Sigma}_q+\nu^2 \mathbf{I}\right)}{\sum_{j=1}^Q     \mathcal{C N}   \!  \left(    \mathbf{v}  ; \boldsymbol{\mu}_j, \boldsymbol{\Sigma}_j    \!   +   \!    \nu^2 \mathbf{I}\right)} .
	\end{aligned}
\end{align}

Using (\ref{mmse_MC}), Equation \eqref{eq:replica-fp} can be calculated by Monte Carlo simulations, which characterizes the Bayes-optimal asymptotic NMSE of any inference scheme for the linear model.
It provides an optimal bound to assess the performance of any channel estimation scheme with the given channel dimension $ M $ and observation dimension $ M_{RF} $ in the assumption that a priori distribution of the channel is known. 


\subsubsection{LS and AE-LS}
As benchmarks, we also introduce the LS solution and the AE-enhanced LS method. We consider the following quadratic criterion
\begin{align}
	\hat{\mathbf{h}}_{k,\mathrm{LS}}
	&= \arg\min_{\mathbf{h}_k}
	\big\|\mathbf{y}_k - \mathbf{V}_{\rm ul} \mathbf{h}_k\big\|_2^2
	+ \lambda_{\mathrm{LS}} \big\|\mathbf{h}_k\big\|_2^2 ,
	\label{eq:ls_opt}
\end{align}
where $\lambda_{\mathrm{LS}}\ge 0$ is a small  regularization parameter to improve the numerical stability of the inversion under   ill-conditioned measurement matrices caused by the sub-connected structure. The closed-form
solution   is given by
\begin{align}
	\hat{\mathbf{h}}_{k,\mathrm{LS}}
	= \big(\mathbf{V}_{\rm ul} ^ {H}\mathbf{V}_{\rm ul} 
	+ \lambda_{\mathrm{LS}}\mathbf{I}_M\big)^{-1}
	\mathbf{V}_{\rm ul} ^{H}\mathbf{y}_k,
	\label{eq:ls_solution}
\end{align}

The LS estimator  does not exploit any structural prior of the near-field channel. To incorporate the learned regularizer in a simple non-iterative manner, we also consider an AE-LS benchmark, which refines the LS estimate by a single pass through the trained regularization operator. Mathematically, we have
$ 	\hat{\mathbf{h}}_{k,\mathrm{AE\text{-}LS}}
= \mathcal{R}_{\boldsymbol{\Theta}^*}\big(\hat{\mathbf{h}}_{k,\mathrm{LS}}\big) $.


\subsubsection{End-to-End Learning}

As another benchmark, we  consider a purely
data-driven approach that directly maps the whitened measurements
$\mathbf{y}_k$ to the corresponding channel vector $\mathbf{h}_k$ via a
deep neural network, referred to as the y2h network $\mathcal{F}_\vartheta(\cdot)$ parameterized
by $\vartheta$. The reconstruction is operated as
$ 	\hat{\mathbf{h}}_{k,\mathrm{y2h}}	= \mathcal{F}_\vartheta(\mathbf{y}_k) $ where real and imaginary parts of $\mathbf{y}_k$ and $\mathbf{h}_k$ are
stacked to the dimension of
$\mathbb{R}^{2M_{RF}}$ and $\mathbb{R}^{2M}$, respectively. Given $  \mathbf{V}_{\rm ul}  $,  we realize
$\mathcal{F}_\vartheta(\cdot)$ as a   fully connected
network trained in a supervised manner as 
\begin{align}
	\vartheta^*
	= \arg\min_{\vartheta}
	\frac{1}{N_{\mathrm{train}}}
	\sum\nolimits_{k=1}^{N_{\mathrm{train}}}
	\big\|
	\mathcal{F}_\vartheta(\mathbf{y}_k) - \mathbf{h}_k
	\big\|_2^2,
	\label{eq:y2h_training}
\end{align}
over a training dataset consisting of $N_{\mathrm{train}}$ channel-measurement pairs ${(\mathbf h_k,\mathbf y_k)}_{k=1}^{N_{\mathrm{train}}}$.

\subsubsection{Learning in the Beamspace}
While the near-field channel is not sparse in the angular domain, the AMP algorithm may be easier to work in the transformation domain $\mathbf{a}_k \triangleq \mathbf{F}\mathbf{h}_k$ where $\mathbf{F}$ is is a unitary DFT matrix
satisfying $\mathbf{F}^H\mathbf{F}=\mathbf{F}\mathbf{F}^H=\mathbf{I}_M$. Accordingly, the observation can be  written as
\begin{align}
\mathbf{y}_k
=
\mathbf{V}_{\rm ul}\mathbf{F}^H\mathbf{a}_k
+\tilde{\mathbf{n}}.
\end{align}

Thus, the channel estimation can be conducted in the DFT domain as
 \begin{align}\label{Lproblem_a}
 		 \hat{\mathbf{a}}_k  \! =	\!    \arg \min _{     \mathbf{a}_k    } \! \! 
 			\left\|     {\mathbf{y}}_{k}   \! -\!    
 			\mathbf{V}_{\rm ul}   \mathbf{F}^H  \mathbf{a}_k 
 			\right\|_2^2 \! +  \!  {\lambda  \!   \left\|     {  \mathbf{a}}_{k}   -
 				\mathcal{R}_{\boldsymbol{\Theta}_a}\left\{  \mathbf{a}_k \right\}
 				\right\|_2^2},
 \end{align}
 where  $\mathcal{R}_{\boldsymbol{\Theta}_a}\{\cdot\}$ is an AE trained using DFT-domain channel samples. This problem can be solved by the same algorithms developed for tackling the antenna-domain estimation problem (\ref{Lproblem}). After obtaining  $ \hat{\mathbf{a}}_k$, we can recover the spatial-domain channel as $\hat{\mathbf{h}}_k= \mathbf{F}^H\hat{\mathbf{a}}_k$.

\section{Transmission Design}\label{section5}
After acquiring CSI $\mathbf{h}_k$, $\forall k$ in each channel coherence time, we focus on the transmission scheme for the XL-MIMO multi-user communication system. This section proposes to jointly design the array geometry and the hybrid precoding to maximize the sum user rate.

Letting $\mathbf{H} = [ \mathbf{h}_1,  \mathbf{h}_2, \ldots,  \mathbf{h}_K ]^H \in\mathbb{C}{^{K \times {M}}}  $ denote the downlink channel from the BS to $K$ users, we can express the downlink received signal at $K$ users as
\begin{align}
	\mathbf{y}  = \mathbf{H}  	\mathbf{x}  + \mathbf{n}  =   \mathbf{H}   \mathbf{V}_{RF}   \mathbf{V}_{BB}      \breve{ \mathbf{s}} + \mathbf{n} ,
\end{align}
where $\mathbf{y}   = [y_1, y_2, \ldots, y_K]^T$ represents the $K$-by-$1$ received signal,  $ \breve{ \mathbf{s}} \in \mathbb{C}^{  {K}}$ is  symbols requested by $K$ users with $\mathbb{E}\left\{ { \breve{ \mathbf{s}} \breve{ \mathbf{s}}^H} \right\} = {\mathbf{I}_K}$, and $\mathbf{n} \sim  \mathcal{CN}\left(\mathbf{0}, \sigma^2 \mathbf{I}_K\right)$ is the thermal noise with variance of $\sigma^2$. 
$\mathbf{V}_{BB}  \in \mathbb{C}^{  M_{RF}\times K} \triangleq [  \mathbf{v}_{B,1},    \mathbf{v}_{B,2},  \ldots,  \mathbf{v}_{B,K}            ] $ and $\mathbf{V}_{RF} \in \mathbb{C}^{ M\times M_{RF}  }$ denote the downlink digital precoder and analog RF precoder at the BS, respectively. Since we consider the sub-connected structure, the RF precoder can be written as $ \mathbf{V}_{RF} \triangleq \mathrm{blkdiag} \left\{       \mathbf{v}_{R,1},    \mathbf{v}_{R,2},  \ldots,  \mathbf{v}_{R, M_{RF}}          \right\} $ with $   \mathbf{v}_{R,m_{RF}}  \in \mathbb{C}^{  {Ms}}$ , $\forall m_{RF}$, whose each element follows   the unit-modulus constraint.

Then,  the  data rate of user $k$ in the investigated multi-user XL-MIMO network can be calculated as
\begin{align}\label{rateExpression}
	 R_k = \log \left(    
	1+ \frac{   \left|     \mathbf{h}_k^H      \mathbf{V}_{RF}    \mathbf{v}_{B,k}      \right|^2  }
	{    \sum_{i=1,i \neq k}^{K} \left|     \mathbf{h}_k^H     \mathbf{V}_{RF}    \mathbf{v}_{B,i }      \right|^2     + \sigma^2           }
	\right).
\end{align}

The  sum user rate maximization problem with hybrid-precoding constraints  and array-geometry constraints can be formulated as follows
\begin{subequations}\label{optimization_problem}
	\begin{align} \label{obj}
		&  \mathop {\max }\limits_{     \{ S, I,\theta, {\mathbf{V}_{{BB}}},{\mathbf{V}_{RF}}   \}    }  \;  \;  \;   {\sum\nolimits_{k = 1}^{{K}} {    \hat{R}_{k}    }}    \\
		\label{power_constraint}
		&\text { s.t. }    \left[     \mathbf{V}_{BB}   \mathbf{V}_{BB} ^H    \right]_{m_{RF} , m_{  RF}   } \le {P  }, \;\; 1\leq m_{  RF} \leq M_{RF}, \\\label{analog_constrinat}
				&       \qquad     \left|       \left[\mathbf{v}_{R,m_{RF}}  \right]_{m_s}        \right|= 1,  \;\;     1\leq m_s\leq M_s, \\\label{curve_constraint}
		&  \qquad    0 \le \theta  \le \pi,  \\\label{size_constraint}
		& \qquad      \left(   {I}      - 1  \right) S d   +   \left(   ({{M_h}/{I}})  -1 \right)     d   \le  L_{h,\max} ,  \\\label{spacing_constraint_h}
		& \qquad       S \ge   {   {M_h}/{I}    } , \\\label{tile_numb}
		& \qquad I \in \mathbb{Z}^+,
	\end{align}
\end{subequations}
where    $ \hat{R}_{k} $ is the data  rate (\ref{rateExpression}) calculated based on the knowledge of channel estimation $\hat{\mathbf{h}}_k$.   
(\ref{power_constraint}) is a practical per-RF chain power constraint which characterizes the fact that   the  power amplified by the $ m_{RF} $-th RF's amplifier  is not allowed to exceed its physical power constraint $P$.
 (\ref{analog_constrinat}) describes the unit-modulus constraint of  phase shifters in the considered sub-connected structure. (\ref{curve_constraint}) describes the curvature region of the cylindrical array which ensures that the array does not bend excessively, avoiding the loss of radiation efficiency towards its coverage sector.   (\ref{size_constraint}) limits the maximal length of the sparse array. (\ref{spacing_constraint_h}) guarantees that subarrays do not overlap with each other.

Problem (\ref{optimization_problem})  is highly non-convex.  To make it tractable, we introduce auxiliary variables ${ {u}_{{k}}}$ and $  {v}_{k} \geq  {0}$  and apply the WMMSE method \cite{shi2011iteratively} for the non-convex objective function (\ref{obj}), leading to the following recast problem
	\begin{subequations}\label{MSE_problem}
	\begin{align}\label{MSE_objective}
		&\min _{  \{ S, I, \theta,  \mathbf{V}_{BB}, \mathbf{V}_{RF},  {v}_{k}, u_k   \}  } \sum\nolimits_{k=1}^K\left(v_k e_k-\log v_k\right) ,\\
		&\qquad {\text { s.t. } } \qquad ({\rm \ref{power_constraint}}) -   ( {\rm \ref{tile_numb}}), \nonumber
	\end{align}
\end{subequations}
where
\begin{align}
	\begin{aligned}
		e_k &= 1+\left|u_k\right|^2        \hat{\mathbf{h}}_k^H \mathbf{V}_{RF}    \mathbf{V}_{BB}  \mathbf{V}_{BB} ^H   \mathbf{V}_{RF}^H\hat{\mathbf{h}}_k\\
	&-2\mathrm{Re}\left\{u_k^H \hat{\mathbf{h}}_k^H \mathbf{V}_{RF} \mathbf{v}_{B,k}\right\}+\sigma^2\left|u_k\right|^2.
	\end{aligned}
\end{align}
which can be solved by  utilizing the block coordinate descent method for variables  $  S$, $I$, $\theta$,  $\mathbf{V}_{BB}$, $\mathbf{V}_{RF}$,  ${v}_{k}$, and $u_k  $.   Note that given other variables, the optimal design of $	 {u}_{k}^{\text {opt}} $ and $ {v}_{k}^{\text {opt}}$ are known after checking the first-order optimality condition of  objective function   (\ref{MSE_objective}), leading to $ v_k^{\rm opt}  =\left(e_k\right)^{-1} $ and
\begin{align}\label{optimal_u}
	u_k^{\rm opt} &=\left(\sum\nolimits_{i=1}^K \hat{\mathbf{h}}_k^H \mathbf{v}_{B,i} \mathbf{v}_{B,i}^H \hat{\mathbf{h}}_k+\sigma^2\right)^{-1} \hat{\mathbf{h}}_k^H \mathbf{v}_{B,k}.
\end{align}

In the following, we will design the array-geometry parameters  $S$, $I$, and $\theta$, the digital precoder $\mathbf{V}_{BB}$, and the analog precoder $\mathbf{V}_{RF}$, respectively.

\subsection{Design of Array Geometry}
In general, the optimal design of array geometry depends on the instantaneous channel realizations and varies with time. However, in practice, the physical architecture of the array  is more likely a pre-decided parameters that should not be adjusted frequently in each channel coherence time. Considering this point, we propose a sub-optimal scheme that decouples the array geometry design from instantaneous channel realizations. Recall that the advantage of modular cylindrical array is to expand the near-field region so that favorable spherical-wave propagation condition is enabled for both the cell center and the cell edge. In other words, it can   provide uniform beam and improve channel condition with near-field effects. Thus, we design the array geometry so that the average channel condition number between multiple users is minimized in the cell. This is because a smaller condition number indicates more orthogonal multi-user channels, a more well-conditioned matrix, and more stable ZF/MMSE precoding. By contrast, a larger condition number indicates more correlated user channels and greater difficulty in spatial separation. Accordingly, the optimization problem is formulated as follows
\begin{subequations}
	\begin{align} \label{1234_objective_funtionss}
		&  \min_{  \{ S, I, \theta \} }  \quad
		\mathbb{E}   \left\{  \text{cond}  \left(     	\bar{\mathbf{H}}    \right)   \right\} , \\
		&\text { s.t. }  \qquad  ({\rm \ref{curve_constraint}}) -   ( {\rm \ref{tile_numb}}),  \nonumber
	\end{align}
\end{subequations}
where $ 	\bar{\mathbf{H}}=
[\bar{\mathbf{h}}_1^{\rm los},\bar{\mathbf{h}}_2^{\rm los},\ldots,\bar{\mathbf{h}}_K^{\rm los}]^H
 $ is the normalized multi-user channel matrix  with
$\bar{\mathbf{h}}_k^{\rm los}=\mathbf{h}_k^{\rm los}/\|\mathbf{h}_k^{\rm los}\|_2$ 
denoting the normalized LoS channel direction of user $k$.
 The expectation is with respect to the random user distribution within the considered service area. This problem can be solved effectively by exhaustive search since all variables are of one-dimensional. 

\subsection{Design of Digital Precoding}
Once the array is designed, we can optimize the digital and analog precoding based on the designed array geometry. To tackle the per-RF chain constraint (\ref{power_constraint}), we define   $ \mathbf{V}_{BB}^H \triangleq \left[ \tilde{\mathbf{v}}_{1}, \tilde{\mathbf{v}}_{2}, \ldots, \tilde{\mathbf{v}}_{M_{RF}}\right] $ so that $\tilde{\mathbf{v}}_{m_{RF}}$ is RF chain-aware variables. Accordingly,  constraint (\ref{power_constraint}) can be rewritten as
\begin{align}\label{per_antenna_cons}
	\begin{aligned}
		 \left[     \mathbf{V}_{BB}   \mathbf{V}_{BB} ^H    \right]_{m_{RF} , m_{  RF}   } = \left\|    \tilde{\mathbf{v}}_{m_{RF}}     \right\|^2 \le P,  \;\;\;\; \forall m_{ RF}.
	\end{aligned}
\end{align}
This allows us to design  alternating optimization algorithms for variables $\tilde{\mathbf{v}}_{m_{RF}}$, $\forall m_{ RF}$. 

Considering the large number of antennas, closed-form solutions with low complexity are highly desired. Thus, after omitting the irrelevant constants, we reformulate the objective function (\ref{MSE_objective}) as follows
\begin{align}\label{function_recast}
	\begin{aligned}
		& \sum\nolimits_{k=1}^K v_k e_k  =    \sum\nolimits_{k=1}^K \operatorname{Tr} \left\{   v_k e_k    \right\}  \\
		& =\operatorname{Tr} \left\{    \mathbf{V}_{BB}  \mathbf{V}_{BB} ^H \sum\nolimits_{k=1}^K  v_k  \left|u_k\right|^2  \mathbf{V}_{RF}^H\hat{\mathbf{h}}_k  \hat{\mathbf{h}}_k^H \mathbf{V}_{RF} \right\} \\
		&- 2  \operatorname{Tr}  \left\{    \operatorname{Re}  \left\{\sum\nolimits_{k=1}^K  v_k u_k^H \mathbf{v}_{B,k }\hat{\mathbf{h}}_k^H  \mathbf{V}_{RF}    \right\}  \right\} \\
		& =\operatorname{Tr}\left\{\mathbf{V}_{BB} ^H  \mathbf{V}_{RF}^H \hat{\mathbf{H}}_{\rm } ^H \mathbf{B} \hat{\mathbf{H}}_{\rm } \mathbf{V}_{RF} \mathbf{V}_{BB} \right\}\\
		&-2 \operatorname{Tr}\{\operatorname{Re}\{\mathbf{C}  \hat{\mathbf{H}}_{\rm }  \mathbf{V}_{RF}  \mathbf{V}_{BB} \}\}
	\end{aligned}
\end{align}
where $\hat{\mathbf{H}} = [ \hat{\mathbf{h}}_1, \hat{ \mathbf{h}}_2, \ldots,  \hat{\mathbf{h}}_K ]^H$ is the overall estimated channel and
\begin{align}
	\begin{aligned}
		& \mathbf{B} \triangleq \operatorname{diag}\left\{v_1\left|u_1\right|^2, \ldots, v_K\left|u_K\right|^2\right\}, \\
		& \mathbf{C}  \triangleq  \operatorname{diag}\left\{v_1 u_1^H, \ldots, v_K u_K^H\right\}.
	\end{aligned}
\end{align}
By defining $  \hat{\mathbf{H}}_{\rm }  \mathbf{V}_{RF}  \triangleq   \left[\tilde{\mathbf{g}}_1, \tilde{\mathbf{g}}_2, \ldots \tilde{\mathbf{g}}_{M_{RF}}\right]  $, we can now explicitly characterize the impact of RF-aware variable  $\tilde{\mathbf{v}}_{m_{RF}}$  through the relationship of 
\begin{align}
 \hat{\mathbf{H}}_{\rm }  \mathbf{V}_{RF} \mathbf{V}_{BB} = \sum\nolimits_{m_{RF}=1}^{M_{RF}} \tilde{\mathbf{g}}_{   m_{RF}   }    \tilde{\mathbf{v}}_{  m_{RF}  }^H.
\end{align}

In an alternating way,  for the $m$-th RF chain, we can arrange the objective function (\ref{function_recast}) with respect to variable $  \tilde{\mathbf{v}}_{m}  $  as follows
\begin{align}\label{conv_func}
	\begin{aligned}
		& f\left(\tilde{\mathbf{v}}_{m^{}}\right) =\mathrm{const} + \left(\tilde{\mathbf{g}}_{m^{}}^H \mathbf{B} \tilde{\mathbf{g}}_{m^{}}\right)\left\|\tilde{\mathbf{v}}_{m^{}}\right\|^2 \\
		&+ \! 2 \operatorname{Re}\!  \left\{  \!   \left(\sum\nolimits_{m_{RF} \neq m }^{M_{RF}} \tilde{\mathbf{g}}_{m^{}}^H \mathbf{B} \tilde{\mathbf{g}}_{m_{RF}} \tilde{\mathbf{v}}_{m_{RF}}^H \! -\!     \tilde{\mathbf{g}}_{m^{}}^H \mathbf{C}^H  \!   \right)  \!  \tilde{\mathbf{v}}_{m} \!  \right\}\\
		& \triangleq g_{m^{}}\left\|\tilde{\mathbf{v}}_{m^{}}\right\|^2+2 \operatorname{Re}\left\{\mathbf{d}_{ m^{}}^H \tilde{\mathbf{v}}_{m^{}}\right\} + \mathrm{const}.
	\end{aligned}
\end{align}

The mimization of quadratic function (\ref{conv_func}) in the constraint of $ \left\|\tilde{\mathbf{v}}_{m^{ }}\right\|^2 \leq P$ is convex.   We can readily obtain the closed-form optimal solution of 
\begin{align} \label{optimal_w}
	\tilde{\mathbf{v}}_{ m^{}}^{\text {opt }}
	=- \min \left\{\frac{1}{g_{m }}, \frac{\sqrt{P}}
	{\left\|    {\mathbf{d}}_{ m^{}   }\right\|_2}\right\} {\mathbf{d}}_{ m^{}}.
\end{align}

\subsection{Design of Analog Precoding}
Since the sub-connected structure is considered for the phase-shifter network, we can rewrite the analog precoder  as $ \mathbf{V}_{RF} = \mathbf{\Psi T}$ where $\mathbf{\Psi}=\mathrm{diag}\left\{  {\psi}_1,{\psi}_2,\cdots,{\psi}_M\right\}$ collects the  design variables for  $M$ phase shifters  and $\mathbf{T}=\mathrm{blkdiag}\left\{ \mathbf{1}_{M_s},\mathbf{1}_{M_s},\cdots,\mathbf{1}_{M_s}\right\}$ describes the sparse sub-connected structure. As a result, the unit-modulus constraint (\ref{analog_constrinat}) becomes $ \left|  {\psi}_m \right| = 1$, $\forall m$.

Again, we aim to obtain low-complexity solutions for XL-MIMO. Recall (\ref{function_recast}) and define vector $\boldsymbol{\psi}=\left[   {\psi}_1,{\psi}_2,\cdots,{\psi}_M\   \right]^T$.  We can reformulate the objective function (\ref{function_recast}) with respect to variable $ \boldsymbol{\psi} $ in the following quadratic form
\begin{align}\label{function_recast_analog}
	\begin{aligned}
		 \sum\nolimits_{k=1}^K v_k e_k   &=\operatorname{Tr}\left\{  \hat{\mathbf{H}}_{\rm } ^H \mathbf{B} \hat{\mathbf{H}}_{\rm }  \mathbf{\Psi}   \mathbf{T} \mathbf{V}_{BB}    \mathbf{V}_{BB} ^H  \mathbf{T}^H \mathbf{\Psi}^H   \right\}\\
		&-2 \operatorname{Tr}\{\operatorname{Re}\{   \mathbf{T} \mathbf{V}_{BB}   \mathbf{C}  \hat{\mathbf{H}}_{\rm }   \mathbf{\Psi}  \}\}\\
		&  =\boldsymbol{    \mathbf{\psi}     }^H     \mathbf{F}_1  \boldsymbol{\mathbf{\psi}}
		-2 \operatorname{Re}\left\{\operatorname{diag}\left\{\mathbf{F}_{2}\right\}^T \boldsymbol{\mathbf{\psi}}\right\} ,
	\end{aligned}
\end{align}
exploiting the properties of $ \operatorname{Tr}\left\{\mathbf{A} {\mathbf{\Psi}} \mathbf{B} {\mathbf{\Psi}}^H\right\}=\boldsymbol{\mathbf{\psi}}^H\left(\mathbf{A} \odot \mathbf{B}^T\right) \boldsymbol{\mathbf{\psi}} $ and 
$  \operatorname{Tr}\{\mathbf{F}_2{\mathbf{\Psi}} \}=\operatorname{diag}\{\mathbf{F}_2\}^T \boldsymbol{\mathbf{\psi}}$ for diagonal matrix $\mathbf{\Psi}$, with definitions of $ \mathbf{F}_1 \triangleq      ( \hat{\mathbf{H}}_{\rm } ^H \mathbf{B} \hat{\mathbf{H}}_{\rm }    )    \odot  (    \mathbf{T} \mathbf{V}_{BB}    \mathbf{V}_{BB} ^H  \mathbf{T}^H    )^T$ and $ \mathbf{F}_{2}  \triangleq   \mathbf{T} \mathbf{V}_{BB}   \mathbf{C}  \hat{\mathbf{H}}_{\rm }  $.

Therefore, the optimization problem for the analog precoder can be reformulated as
\begin{subequations}\label{optimization_HMIMO_phase}
	\begin{align}\label{30a}
		&\min\limits_{    \boldsymbol{\psi} }  \;\;\;   f_{0}\left(   {\boldsymbol{\psi}}   \right) = \boldsymbol{    \mathbf{\psi}     }^H     \mathbf{F}_1  \boldsymbol{\mathbf{\psi}}
		-2 \operatorname{Re}\left\{\operatorname{diag}\left\{\mathbf{F}_{2}\right\}^T \boldsymbol{\mathbf{\psi}}\right\} 
		\\\label{constraint_HMIMO_phase}
		&\text { s.t. } \quad    \left|   {   {\psi}}  _m    \right| =1, \;\;  1 \leq m \leq M,
	\end{align}
\end{subequations}
which can be solved iteratively based on the maximization-minorization (MM) algorithm with a closed-form solution in each iteration. Specifically, employing inequality \cite[(61)]{zhi2022ZF}, we have the following surrogate function given a fixed point $ {\boldsymbol{\psi}}_{0}$:
\begin{align}\label{surrogate_dunction}
	f_{0}\left(     {\boldsymbol{\psi}}      \right) \leq - 2 \operatorname{Re}\left\{  \overline{\mathbf{w}}^H     {\boldsymbol{\psi}}       \right\}+ \mathrm{const}
\end{align}
where
\begin{align}
	\overline{\mathbf{w}}^H=  {     {\boldsymbol{\psi}}     }_0{ }^H\left(
	\lambda_{\max }\left\{\mathbf{F}_1\right\} \mathbf{I}_M
	\! -\!  \mathbf{F}_1
	\right)^H
	+\mathrm{diag}\left\{\mathbf{F}_2\right\}^T .
\end{align}
Under constraint (\ref{constraint_HMIMO_phase}), it is readily found that the minimization of surrogate function  (\ref{surrogate_dunction}) has the  optimal solution of
\begin{align}\label{optimal_solution}
	{\boldsymbol{\psi}}^{\rm opt}  = \underset{  {\boldsymbol{\psi}}    }{\arg \min }\left\{-2\operatorname{Re}\left\{   \overline{\mathbf{w}}^H {{\boldsymbol{\psi}}  }   \right\}    \right\}=\exp \{j \angle \overline{\mathbf{w}}     \},
\end{align}
where $[ \overline{\mathbf{w}}]_m\triangleq |  \overline{w}_m   | e^{j \angle  \overline{w}_m}$, $\forall m$, and $\angle{ \overline{\mathbf{w}}     }=[\angle  \overline{w}_1, \cdots, \angle  \overline{w}_M]^T$. 

Following the idea of MM algorithm, Problem (\ref{optimization_HMIMO_phase}) can be effectively solved by  minimizing its upper-bounded  surrogate function (\ref{surrogate_dunction}) with solution (\ref{optimal_solution}) and updating the value of fixed  point by ${\boldsymbol{\psi}}^{\rm opt}  \to  {\boldsymbol{\psi}}_{0} $,  iteratively, until the convergence.


\section{Simulation}\label{section6}
Unless otherwise stated, we consider a BS with $M=48\times16=768$ antennas located at $[0,0,0]$ where each RF chain connects $M_s=6$ antennas, operating at $f_c=6.8$ GHz to communicate with $K$=10 users.   For gradient descent and PnP algorithms, we set a grid of $[10^{-4},10^{-3},\cdots,10^4]$ for  searching the best value of $\lambda$ and $\mu$.   
For the neural network model, the encoder contains three 1D convolutional layers with a kernel size of $5$ and a stride of $2$.
The dimension of the latent vector is $256$. The decoder mirrors the encoder with  three transposed convolutional layers. For offline training,
we generate $25\times 10^4$ channel samples for UPA and MCA where each location of the user  is sampled  randomly from  a 3D area spanning $x\in[5,50]$, $y\in  [-50,50]$, and $z\in[0,4]$, associated with $1-3$ NLoS scatterers randomly sampled from  four small 3D regions of size $5\times5\times2$ located around the 3D region. Besides, each subarray with $6\times4$ antennas is a VR region with a visible probability of $0.8$ for each path. $80\%$ channel data is used for training while the other $20\%$ is used for testing.  
The batch size, number of epochs, and initial learning rate are set to $512$, $100$,
and $10^{-3}$, respectively. The noisy inputs for AE training are generated with SNR uniformly sampled from $0$ dB to $20$ dB.
$\tau=3K$ pilots are sent to improve the number of observations. Normalized mean square
error (NMSE) $ \frac{\|\hat{\bm h}-\bm h\|_2^2}{\|\bm h\|_2^2} $  and normalized correlation coefficient $ \rho = \frac{|\hat{\bm h}^{H}\bm h|}
{\|\hat{\bm h}\|_2\|\bm h\|_2} $ are used to evaluate the performance of channel estimation algorithms

\subsection{Comparing Estimation Algorithms for UPA}

 \begin{figure}[t]
 	\centering
 	\includegraphics[width= 0.4\textwidth]{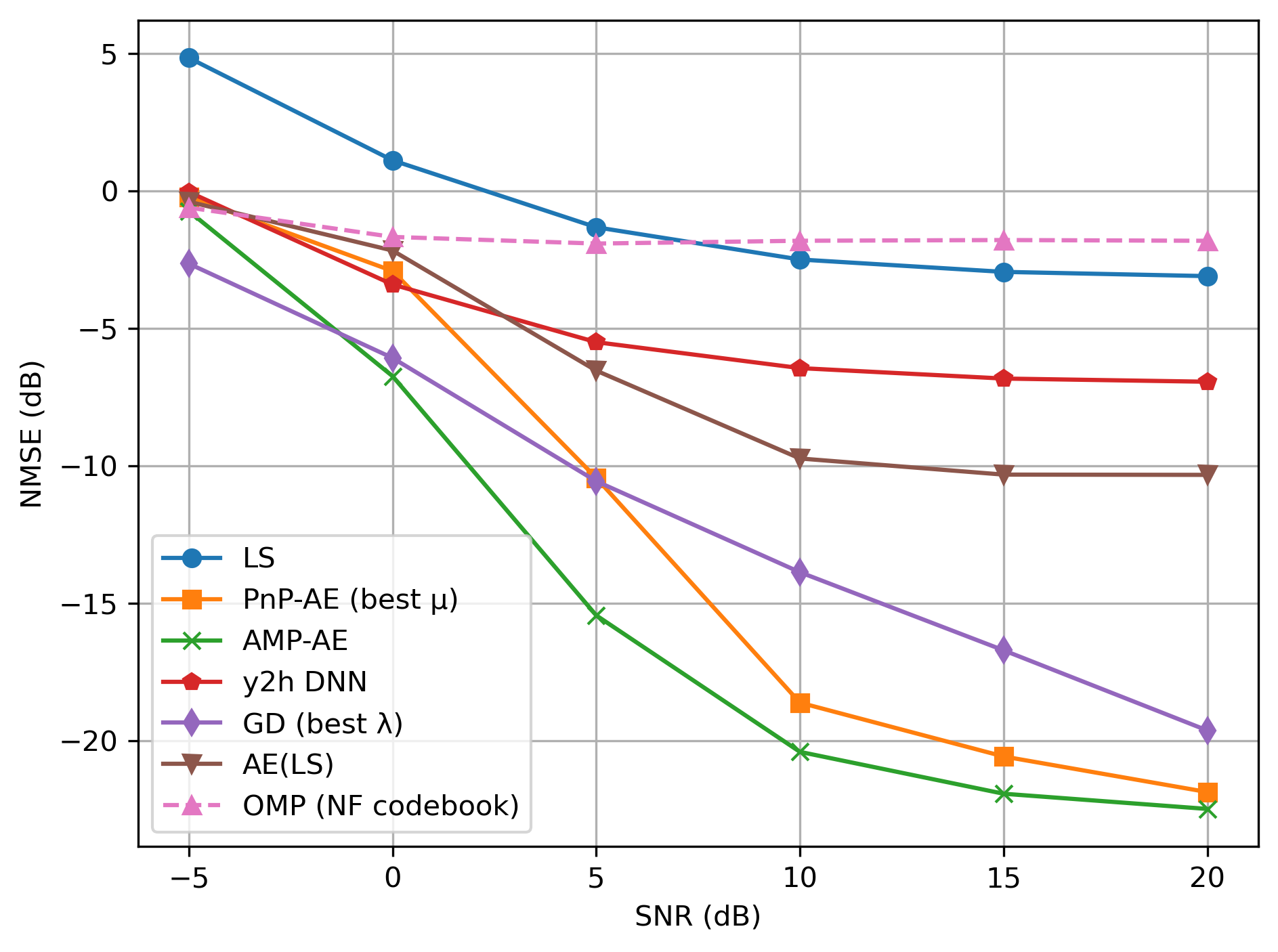}
 	\DeclareGraphicsExtensions.
 	\caption{NMSE performance versus uplink SNR (effective uplink SNR including the channel power gain, same for the subsequent figures).}
 	\label{figure3}
 \end{figure}

 \begin{figure}[t]
 	\centering
 	\includegraphics[width=0.4\textwidth]{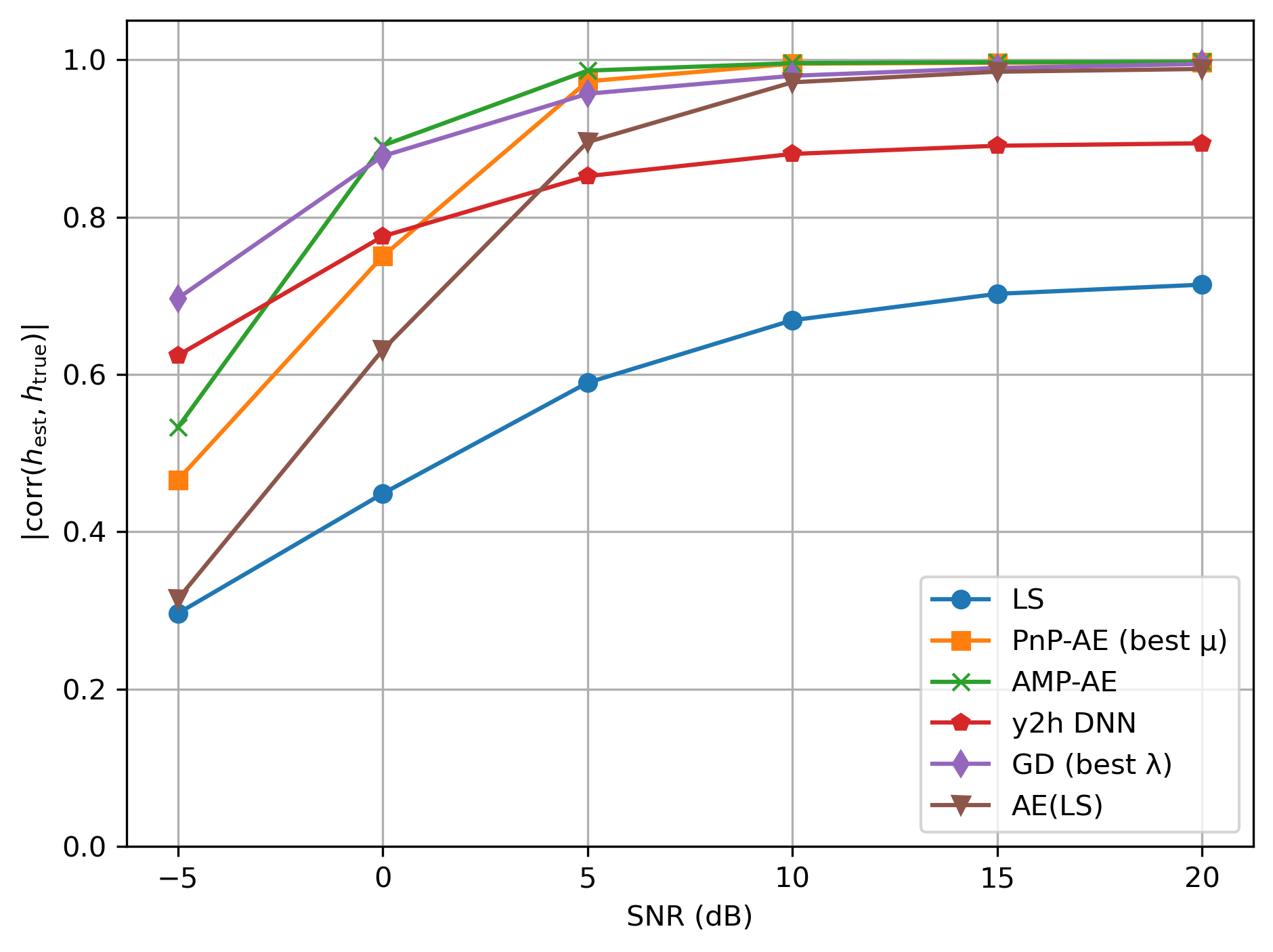}
 	\caption{Correlation coefficient between estimated and true channels.}
 	\label{figure4}
 \end{figure}
 
 We first evaluate the quality of estimated channels  under UPA structures.  In Fig. \ref{figure3}, we compare the proposed AE-AMP algorithms with other benchmarks in terms of NMSE performance. It can be seen that the proposed AE-based AMP algorithm achieves the best performance except for the smallest SNR, showing obvious gains compared to conventional codebook-based algorithms\cite{wu2023multiple} and end-to-end learning methods. This is because conventional algorithms suffer from the complicated near-field channel features, making the estimation problem highly non-trivial given a limited-size codebook. Also, end-to-end learning methods face the very challenging problem of estimating a high-dimensional spatial-domain channel from limited observations at the BS with a sub-connected structure, which limits their performance. Besides, the  y2h network has to be trained for a specific realization of measurement matrix $\mathbf{V}_{\rm ul} $ which loses universality  to practical application. By contrast, the proposed AE-AMP algorithm only uses a network to learn the low-dimensional channel manifold and assists the compressed sensing algorithm with prior information, which is very effective and does not rely on a specific RF combining matrix/measurement matrix. It is worth noting that by exploiting the reconstruction capability of the AE, the proposed denoising network does not require a separate model to be trained for each SNR level, thereby improving its robustness and applicability across a wide range of practical scenarios.  Furthermore, while gradient descent and AE-PnP can also achieve good performance, they rely on the search of weight factors on the predefined grid, increasing the computational complexity and reducing the robustness. Overall, this demonstrates the promising advantages of the proposed AE-AMP algorithm, which is general, effective, and robust.
 
Fig. \ref{figure4} illustrates the similarity of the estimated channel with the true channel. It can be seen that the proposed AE-AMP algorithm quickly achieves a correlation coefficient of $1$ with true channels. This demonstrates that the estimated channels almost have the same spatial structure as the true channel and thus can be used reliably for the   precoding design in downlink transmission.

\begin{figure}[t]
	\centering
	\begin{minipage}{0.65\linewidth}
		\centering
		\includegraphics[width=\linewidth]{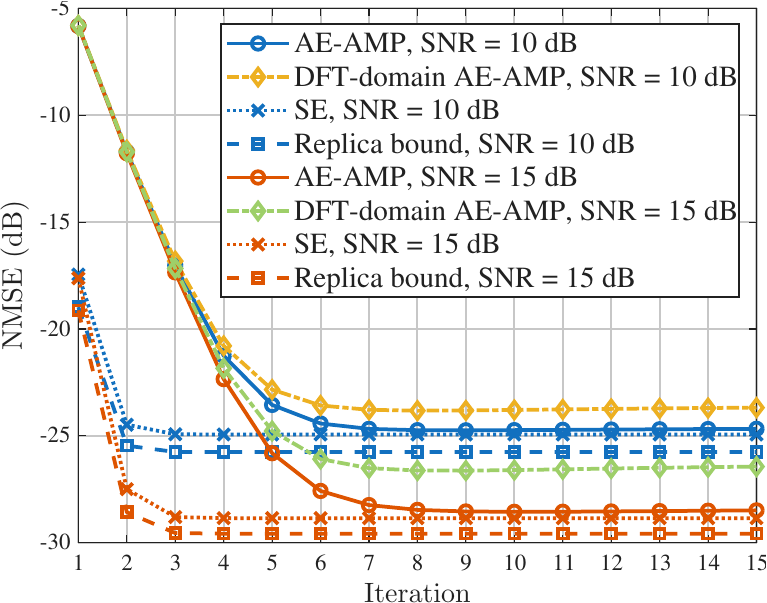}
		\centerline{\footnotesize (a)}
	\end{minipage}
 
	\begin{minipage}{0.65\linewidth}
		\centering
		\includegraphics[width=\linewidth]{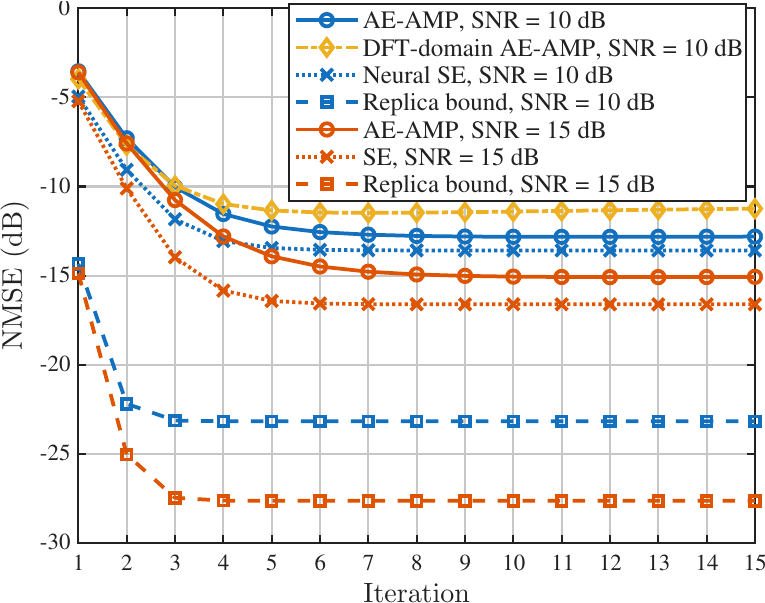}
		\centerline{\footnotesize (b) }
	\end{minipage}
	\caption{Convergence comparison among AE-AMP, DFT-domain AE-AMP, neural state evolution, and the replica bound: (a)  $Q=1$; (b) $Q=10^6$.}
	\label{figure32}
\end{figure}

Fig. \ref{figure32} evaluates the performance of the proposed AE-AMP algorithm with the theoretical bounds (\ref{eq:replica-fp}). We discretize the continuous large-scale channel space into $Q$ states so that the ideal posterior mean estimator can be calculated as (\ref{mmse_MC}). This allows us to assess how closely the proposed estimator approaches the theoretical optimality. We consider a challenging scenario with $L=20$ paths and with strong NLoS components, where the NLoS pathloss is determined by the sum of the two reflected path lengths rather than their product. Firstly, it is observed that the proposed AE-AMP curves closely follow the SE predictions \eqref{eq:se}  over different SNR values, as expected for AMP algorithms. The small gap can be attributed to several practical non-idealities, including the non-i.i.d. feature of the measurement matrix. Secondly, when $Q=1$, i.e., the network only needs to fit a single statistical distribution $p(	\mathbf h_k\mid \boldsymbol{\zeta}_q )$, which is simple. It can be seen that our method closely approaches the replica bound, although the network is trained over a range of SNR values and operates under non-idealities of sub-connect combining matrix, finite-dimensional observations, damping, and non-uniform pathloss. As $Q$ increases, the AE-AMP algorithm experiences some performance loss. This is because our network does not know exact statistical prior and must learn across $Q$ distinct large-scale channel states, which is very difficult as the prior distribution becomes involved. Nevertheless, our scheme still achieves competitive performance, and outperforms its DFT-domain counterpart. This is because the near-field channel is no longer sparse in the angular domain, especially with a large number of paths. Learning directly in the spatial domain therefore avoids reliance on a predefined near-field codebook and is better suited to the rich-scattering near-field environment.

\subsection{Impacts of Adjusting Array Geometries}

\begin{figure}[t]
	\centering
	\includegraphics[width=0.45\textwidth]{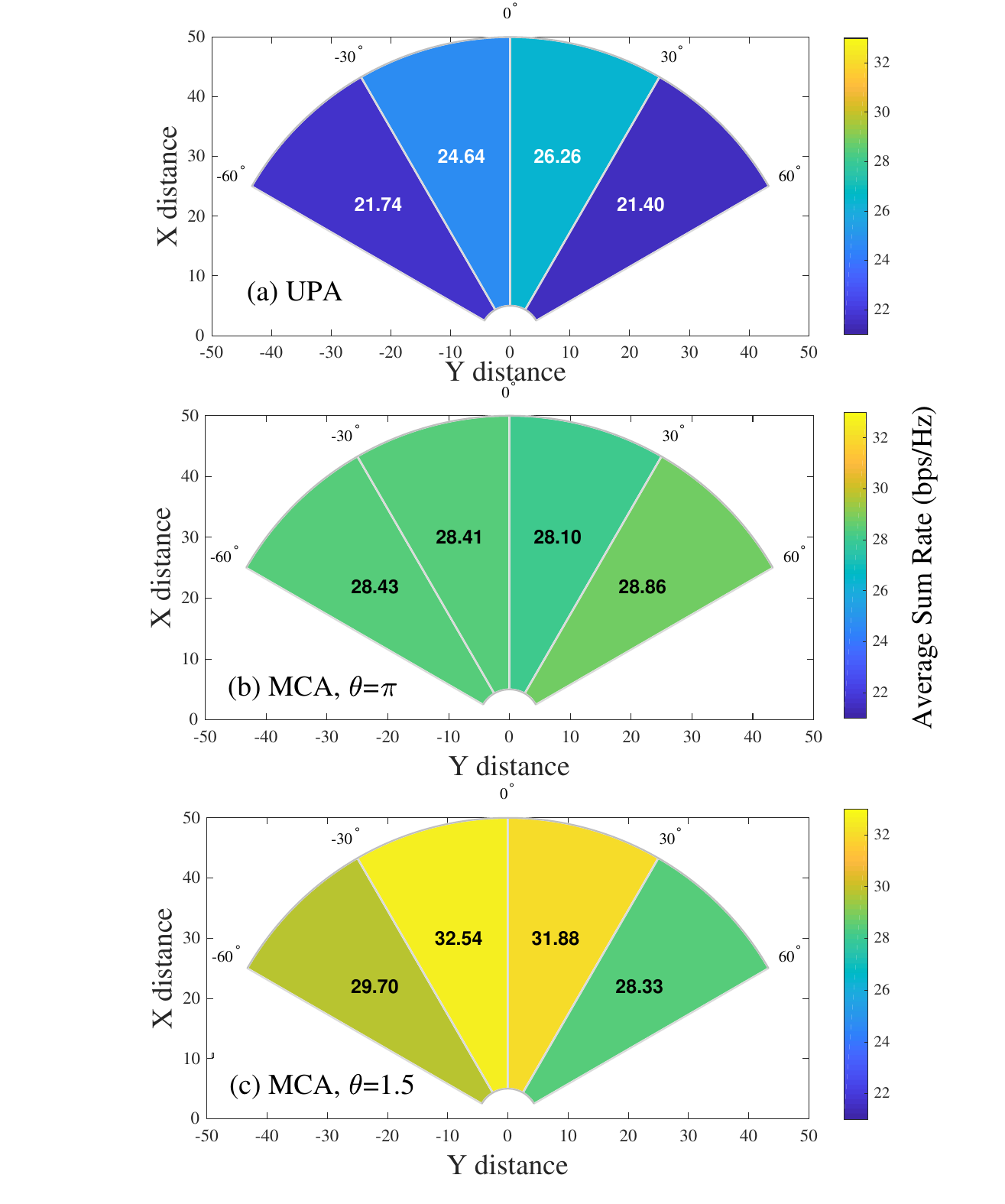}
	\caption{Heatmap of sum data rates for users  located in different $30^{\circ}$ sectors, where $z=0$, $M=12\times12$, $I=4$, $S=10\times\frac{M_h}{I}$.}
				\vspace{-10pt}
	\label{figure14_2}
\end{figure}

Next, we showcase the rate performance for users located in sectors with different directions in the presence of true LoS channels. It can be seen from Fig. \ref{figure14_2} (a) that due to the smaller near-field region in the side directions, the rate heatmap of UPA has some performance loss in the cell edge. By contrast, with the help of uniform beams provided by curved array with a large curvature of $\theta=\pi$, Fig. \ref{figure14_2} (b) illustrates the capability of bending array to achieve a better coverage for the whole cell. Furthermore, by controlling the curvature angles to sophisticatedly improve the channel quality, it can be seen from Fig. \ref{figure14_2} (c) that MCA can enhance the performance not only at the cell edge, but also at the cell center.

\subsection{Rate Comparisons with Estimated Channels}

\begin{figure}[t]
	\centering
	\includegraphics[width= 0.4\textwidth]{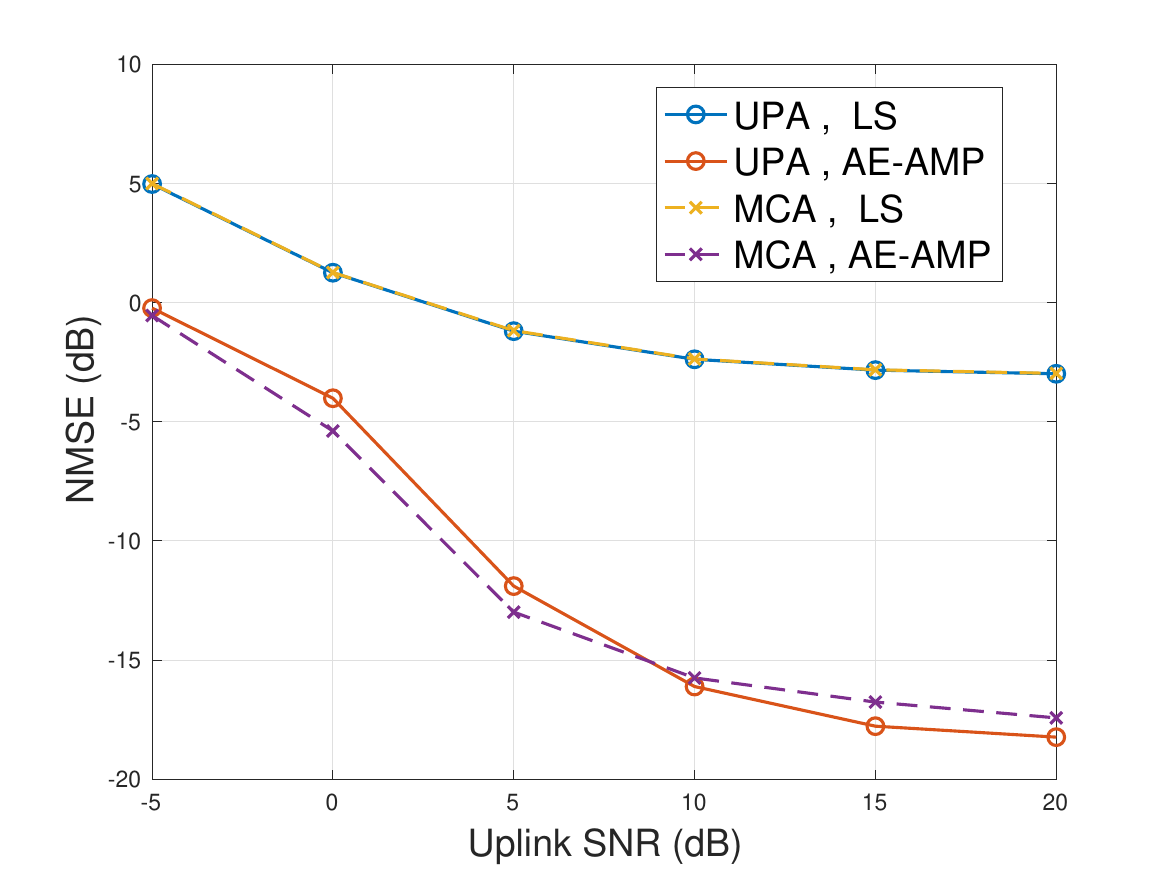}
	\DeclareGraphicsExtensions.
				\vspace{-10pt}
	\caption{NMSE performance for estimating channels of UPA and MCA.}
				\vspace{-15pt}
	\label{figure15}
\end{figure}

\begin{figure}[t]
	\centering
	\includegraphics[width= 0.4\textwidth]{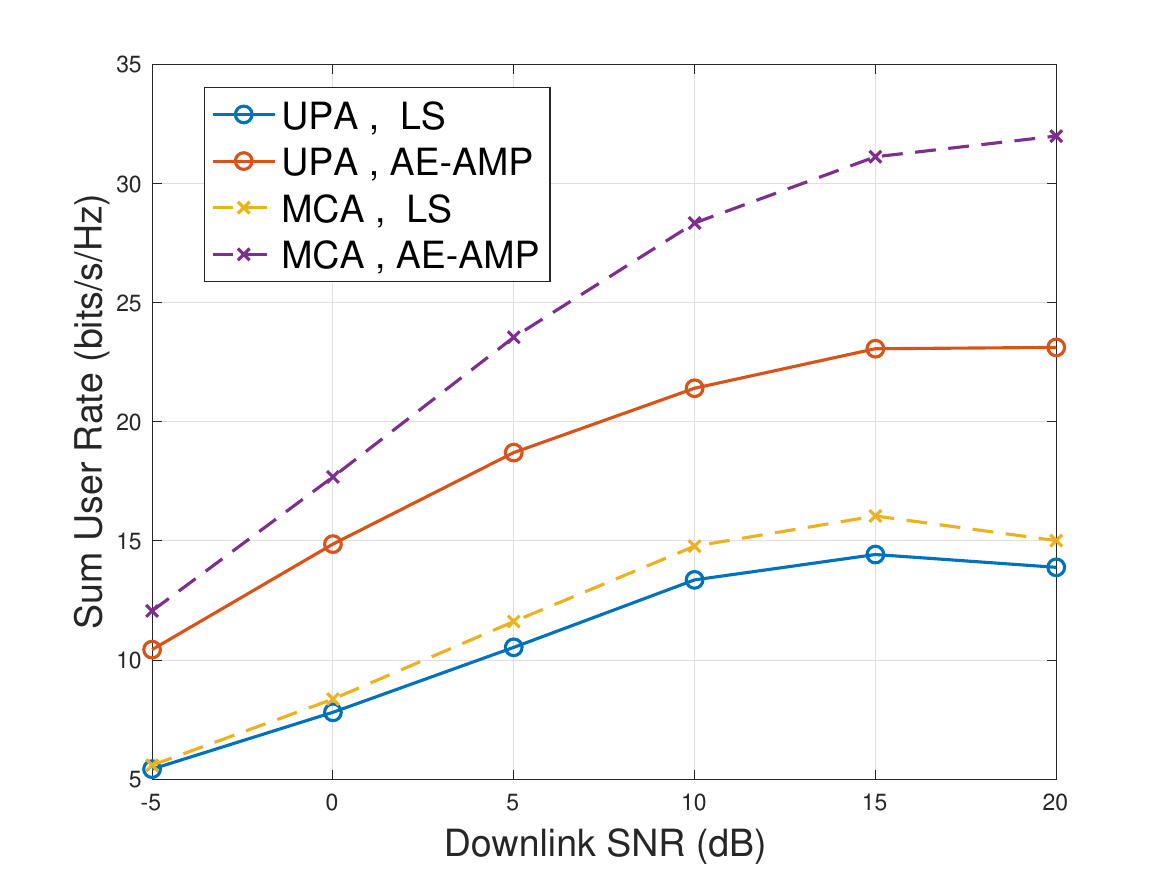}
\vspace{-10pt}
	\caption{Rate comparison    between UPA and MCA with estimated channels.}
			\vspace{-15pt}
	\label{figure22}
\end{figure}
Finally, we evaluate the rate performance for UPA and designed MCA based on the CSI estimated from the proposed algorithms. We consider users to be randomly located in a sector ranging from $10^{\circ}$ to $40^{\circ}$ and evaluate with $200$ random realizations.  In Fig. \ref{figure15}, we first illustrate the NMSE performance of the channel estimation algorithms for UPA and MCA. It can be seen that compared to the LS method, the proposed AE-AMP algorithm     maintains high-accuracy NMSE performance for different array geometries, demonstrating its generality with different array geometries. The similar estimation precision achieved between UPA and MCA also shows the capability of the algorithm to learn the channel manifold even with more complicated geometries.

  Fig. \ref{figure22} illustrates the downlink achievable rate performance  with CSI estimated at $0$ dB uplink SNR. It can be seen that with the highly accurate CSI,  precoding design based on channels estimated by AE-AMP algorithms achieves an obvious gain compared to that with LS channel estimation. This supports the reliability of the proposed channel estimation algorithms in precoding design. Besides, the rate gain brought by optimized array geometries is also demonstrated compared to UPA, due to the better channel condition and larger near-field region achieved by MCA. The main reason is that MCA has a better beamfocusing capability to mitigate multi-user interference even when users are located with closed angles.

\section{Conclusion}\label{section7}
	This paper investigated the challenging problem of XL-MIMO communication with  array geometry optimization and CSI acquisition. We first established the  channel model with different array geometries and analysed the near-field effects. Then, we proposed a general deep learning-based channel estimation algorithm that employs the learned prior information to support compressed sensing methods. Finally, we proposed the optimization algorithms for designing array topologies and hybrid precoding in the downlink. Numerical results demonstrated its superiority over several benchmarks.

\bibliographystyle{IEEEtran}
\bibliography{myref}

\end{document}